\newcommand{\BibitemShut}[1]{}
\documentclass[aps,twocolumn,superscriptaddress]{revtex4-1}

\usepackage{latexsym}
\usepackage{graphicx}
\usepackage{color}

\begin{document}



\title{The question of intrinsic origin of the metal-insulator transition in i-AlPdRe quasicrystal}


\author{Julien Delahaye}
\affiliation{Institut N\'eel, Universit\'e Grenoble Alpes and CNRS, BP 166, F-38042 Grenoble C\'edex 9, France}
\author{Claire Berger}
\affiliation{Institut N\'eel, Universit\'e Grenoble Alpes and CNRS, BP 166, F-38042 Grenoble C\'edex 9, France}
\affiliation{School of Physics, Georgia Institute of Technology, Atlanta, GA 30332, USA}


\date{\today}

\begin{abstract}

The icosahedral (i-) AlPdRe is the most resistive quasicrystalline alloy discovered so far. Resistivities ($\rho$) of $1\Omega cm$ at 4K and correlated resistance ratios ($RRR = \rho_{4K}/\rho_{300K}$) of more than 200 are observed in polycrystalline samples. These values are two orders of magnitude larger than for the isomorphous i-AlPdMn phase. We discuss here the controversial microscopic origin of the i-AlPdRe alloy electrical specificity.  It has been proposed that the high resistivity values are due to extrinsic parameters, such as secondary phases or oxygen contamination. From comprehensive measurements and data from the literature including electronic transport correlated with micro structural and micro chemical analysis,  we show that on the contrary  there is mounting evidence in support of  an origin intrinsic to the i-phase. Similarly to  the other quasicrystalline alloys, the electrical resistivity of the i-AlPdRe samples depends critically on  minute changes in the structural quality and chemical composition. The low resistivity in i-AlPdRe single-grains compared to polycrystaline samples can be explained by difference in chemical composition,  heterogeneity and thermal treatment.
\end{abstract}



\maketitle


\section{\label{Introduction}Introduction}

The icosahedral (i-) AlPdRe alloy has unique electronic properties compared to all the other quasicrystalline alloys discovered so far. Specifically, some of the more resistive i-AlPdRe polycrystalline samples lie on the insulating side of the metal-insulator transition (MIT) \footnote{An insulator is by definition a material with a zero conductivity (infinite resistivity) at zero temperature, while the conductivity remains non-zero (finite resistivity) in a metallic system.}, with conductivity  at 4K as low as $\sigma_{4K}=1\Omega^{-1}cm^{-1}$ and resistance ratios between 4K and 300K ($RRR=\rho_{4K}/\rho_{300K}$) larger than 200 \cite{RappBook99, PoonMRS99, DelahayeJPCM03}. In other words, this alloy of metals becomes an insulator, the origin for this remarkable result being attributed to the specificity of the i-AlPdRe phase. However, in the early 2000's, electrical measurements  performed on i-AlPdRe single-grains \cite{GuoPML00, FisherPMB02} showed significantly higher conductivity $\sigma_{4K}>200\Omega^{-1}cm^{-1}$ and lower $RRR < 2.5$, thereby placing i-AlPdRe single grains on the metallic side of the MIT. These values are similar to those observed in the i-AlCuFe and i-AlPdMn phases. A controversy  followed as to wether the high electrical resistivities observed in some i-AlPdRe polycrystalline samples are intrinsic to the i-phase or due to extrinsic effects, such as impurities, secondary phases or grain boundaries.

Extrinsic origin proponents  argued \cite{DolinsekPRB06}  that the high resistivities of  polycrystalline i-AlPdRe ingots result from a combination of a high porosity and the existence of oxygen-rich insulating regions in the material. A granular model was further proposed  \cite{VekilovEPL09}  to describe the electronic properties of the i-AlPdRe polycrystalline samples. Counter-arguments  were raised later \cite{PoonPRB07,RappPRB11}, based on the similarity of the temperature and magnetic field dependence of the conductivity $\sigma(T,H)$ for  polycrystalline ribbons and ingots, in spite of their different morphology. 
The main point of the controversy \cite{DolinsekPRB07} is that there is so far no physical picture which includes all the i-AlPdRe samples, i.e. single- and poly-grains.

In this paper, we comment on these recent publications and include unpublished results on a large number of melt-spun ribbons \cite{DelahayeThesis99}. We give a comprehensive overview of  all types of i-AlPdRe samples (single-grains, polycrystalline ingots, ribbons and thin films) from the literature and our own work. In particular, quantitative chemical composition analysis of ribbons and ingots shed some new light on the origin of high resistivity values. Our main point can be summarized as follows.

\begin{enumerate}

     \item We provide arguments why the case for an extrinsic origin of the insulating behavior in i-AlPdRe samples \cite{FisherPMB02,DolinsekPRB06,DolinsekPRB07} is inconclusive. Specifically,  our chemical analysis shows that the composition range common to highly resistive ribbons and ingots is significantly different from that of the few metallic single-grains measured. Detailed investigation of existing data disproves the granular electronic model \cite{VekilovEPL09}, the role of porosity and questions the oxidation effect.

 \item Despite difference in their microstructure, ribbons and ingots show insulating behavior, in agreement with previous report by Poon and Rapp  \cite{RappPRB11}. The presence of defects and chemical inhomogeneity influences the electrical properties (especially at low temperature), but cannot explain the difference between metallic and insulating samples.

   \end{enumerate}


\section{\label{AlPdReSamples}The different types of i-AlPdRe samples}

Different types of i-AlPdRe samples have been produced (see Table \ref{TableSamples} and Ref. therein): single-grains, polycrystalline ingots, ribbons and thin films. Note that the thin films have not been included in the recent discussions about the origin of the high resistivity values.

Fabrication of ingots \cite{HondaJJAP94,PiercePRL94,SawadaICQ6,RosenbaumJPCM04,BianchiPRB97} and ribbons \cite{HondaJJAP94,BergerSSC93,DelahayeDEA97} starts with arc-melting an ingot, containing the desired proportion of Al, Pd and Re, as homogeneously as possible. The ingot then can  be cut in small bars of typical size $1mm\times 1mm\times 5mm$ (called ``ingots''), or melt-spun into ribbons of typical size $20\mu m \times 1mm \times 5mm$.

Thin films,  typically 200nm thick, were made by using either co-evaporation \cite{HaberkernCondmat99} or  sequential evaporation techniques \cite{BergmanStage99}. With the co-evaporation technique, Al-Pd and Re sources are used, and depending on the substrate position during  evaporation, different Re content can be achieved. With the sequential evaporation technique, layers of pure elements are evaporated sequentially and the relative thickness of  each layer determines the chemical composition of the film.
For all these polycrystalline samples, a subsequent annealing is necessary to get a sample of high structural quality  and high resistivity values (see  subsection \ref{Annealing}).

Single-grains of millimetre size were grown by a slow cooling technique starting from a homogeneous melt of compositions close to the icosahedral phase domain \cite{GuoPML00,FisherPMB02}. The single-grains can then be separated from the remaining melt by decanting at low temperature.

\begin{table*}[htb]
\caption{\label{TableSamples}
i-AlPdRe samples and their measured  electrical resistivity. From  left to right: first year of publication by the research group, reference name of the group,  type of samples, nominal (or * measured) composition giving the highest resistivities,  maximum RRR reached and the related reference.}
\begin{ruledtabular}
\begin{tabular}{|c|c|c|c|c|c|}
  \hline
  1993 & Takeuchi et al & Ingots and ribbons & $Al_{70}Pd_{20}Re_{10}$ & 51 & \cite{HondaJJAP94}\\
  1993 & Poon et al & Ingots & $Al_{70.5}Pd_{21}Re_{8.5}$ & 280 & \cite{PoonMRS99}  \\
  1993 & Berger et al & Ribbons & $Al_{70.5}Pd_{21}Re_{8.5}$ & 209 & \cite{DelahayeJPCM03} \\
  1995 & Kimura et al & Ingots & $Al_{70.5}Pd_{22}Re_{7.5}$ & $>$ 10 & \cite{SawadaICQ6} \\
  1996 & Lin et al & Ingots & $Al_{70}Pd_{22.5}Re_{7.5}$ & 136 & \cite{RosenbaumJPCM04} \\
  1996 & Chernikov et al & Ingots & $Al_{70}Pd_{21.4}Re_{8.6}$ & $\simeq$ 10 & \cite{BianchiPRB97} \\
  1997 & Calvayrac et al & Ribbons & $Al_{70.5}Pd_{21}Re_{8.5}$ & 130 & \cite{DelahayeDEA97} \\
  1997 & Haberkern et al & Thin films & $Al_{72.3}Pd_{20.2}Re_{7.5}$ & $\simeq$ 20 & \cite{HaberkernCondmat99} \\
  1998 & Grenet et al & Thin films & $Al_{69}Pd_{22}Re_{9.4}$ & 4 & \cite{BergmanStage99} \\
  2000 & Tsai et al & Single-grains & $Al_{71.7}Pd_{19.4}Re_{8.9}*$ & 1.8 & \cite{GuoPML00} \\
  2002 & Fisher et al & Single-grains & Unknown & 2.5 & \cite{FisherPMB02} \\
  \hline
\end{tabular}
\end{ruledtabular}
\end{table*}

\section{\label{SingleGrainsResults} Origin of the high resistivity values:  single-grain results}

 Based on their single-grain measurements, Fisher et al \cite{FisherPMB02} and Dolin\v sek et al \cite{DolinsekPRB06,DolinsekPRB07} have suggested that the high resistivity values observed in some i-AlPdRe polycrystalline samples are extrinsic to the i-phase. In this section, we critically discuss the different arguments leading to this conclusion, which are the following:
 \begin{itemize}

     \item Argument 1: the i-AlPdRe single-grains show no evidence of a metal-insulator transition as a function of either composition or structural perfection. Therefore the conductivity mechanism of the i-AlPdRe phase cannot be dramatically different from other quasicrystal families.

    \item Argument 2: The i-AlPdRe ingots have voids and oxygen rich bridges which are responsible for their high resistivity values.

    \item Argument 3: the density of states (DOS) of AlPdRe and AlPdMn are only marginally different so that electrical transport should be the same.

    \item Argument 4: the high resistivity values come from secondary phases produced by an annealing at too high temperature.

   \end{itemize}

 Some of the criticisms developed here were already mentioned by Poon and Rapp \cite{PoonPRB07}.

\subsection{Argument 1: no evidence of a metal-insulator transition in i-AlPdRe single-grains}

Clearly, all the i-AlPdRe single-grains that have been made so far lie on the metallic side of the metal-insulator transition, with resistivity values similar to what is observed in high quality i-AlPdMn and i-AlCuFe samples. This however doesn't imply that i-AlPdRe samples cannot have larger intrinsic resistivity values.

\subsubsection{Single-grains versus polycrystalline samples: the chemical composition}

For a relevant discussion of the  composition domains of single-grains and polycrystalline samples, it is necessary to compare the actual compositions of the final samples (after annealing) and not the nominal ones. Such a comparison is difficult for at least two reasons. First, the actual chemical composition of i-AlPdRe samples is not necessarily known, and second significant spatial inhomogeneities are often present. According to our microprobe analysis performed on some ribbons and ingots \cite{DelahayeThesis99} (see subsection \ref{CompositionOrder}), the high resistivity values are observed for 66-70 atomic (at.) \%  Al, 21-26 at.\%  Pd and 7-10 at.\%  Re (the nominal composition was $Al_{70.5}Pd_{21}Re_{8.5}$). In single-grains, Guo et al \cite{GuoPML00} have measured a RRR of 1.8 ($\sigma_{300K}\simeq 300\Omega^{-1}cm^{-1}$) for a chemical composition of $Al_{71.7\pm0.9}Pd_{19.4\pm1.6}Re_{8.9\pm0.9}$. Fisher et al \cite{FisherPMB02} have measured the electrical resistivity of six single-grains with different compositions and found a maximum RRR value of 2.5 ($\sigma_{300K} \simeq 300\Omega^{-1}cm^{-1}$). The  composition of  three single-grains  were determined to be in the range 71.6-73.5 at.\% Al, 17.1-19.6 at.\%Pd and 8.8-9.4 at.\%Re. The compositions of the two most resistive single grains are not known, but interestingly enough, it can be guessed from  Table 1 of Ref. \cite{FisherPMB02} that these two single-grains are the closest to the composition domain given above for highly resistive ribbons and ingots. Thus it seems that all the i-AlPdRe single-grains made so far have a chemical composition different from highly resistive ribbons and ingots, with slightly more Al and less Pd, the typical difference being of a few at.\%. Another hint of a composition difference between the single-grains and the ribbons is the fact \cite{FisherPMB02} that  annealing  the single-grains at $1000^oC$ results in a partial decomposition while no decomposition is observed in ribbons annealed up to $1020^oC$ (see Appendix \ref{AppendixA}).

 Can a difference in composition of a few at.\% in Al, Pd or Re  change the  RRR from 2 to more than 200, i.e.  increase the RRR by two orders of magnitude? From the Pd dependence of the RRR plotted in  Figure 8 of Ref. \cite{FisherPMB02} a change of about 3 at.\% in Pd  increased the RRR by less than a factor of two, from $\simeq 1.3$ to $\simeq 1.8$, which lead to the author's conclusion  that \cite{FisherPMB02}:``there is no evidence that this variation in the resistivity with changes in composition could lead to RRR values as large as those observed for some quenched and annealed samples''. This conclusion is based on a linear extrapolation of the conductance dependence on the driving parameter such as the chemical composition. Such linear behavior would be very unusual  close to the MIT. In doped semi-conductors, while the room temperature conductivity (or resistivity) changes slowly with the dopant concentration, the RRR varies much more rapidly across the MIT. For example, in Si-As \cite{CastnerPRB88}, a change of the room temperature resistivity from $7.7 m\Omega cm$ to $8.7 m\Omega cm$ corresponds to an increase of  RRR from 4 to 300 (see Table 1 of Ref. \cite{CastnerPRB88}).

\begin{figure}[h]
\includegraphics[width=8cm]{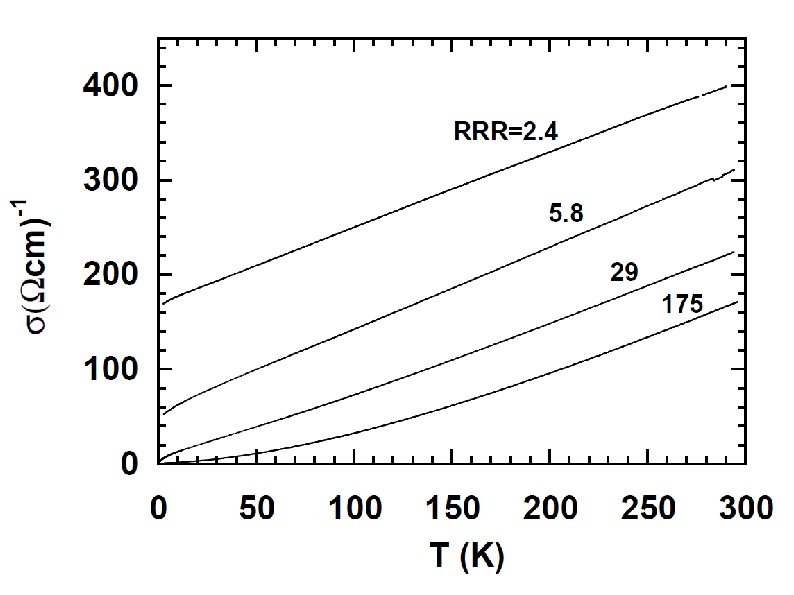}
\caption{\label{Figure1}Conductivity versus temperature for i-AlPdRe melt-spun ribbons of different RRR \cite{DelahayeJPCM03,DelahayeThesis99,DelahayePRB03,DelahayePRB01}.}
\end{figure}

The same idea is illustrated in Figure \ref{Figure1}, where the conductivity of i-AlPdRe melt-spun ribbons with RRR between 2.4 and 175 is plotted below 300K. It is clearly observed  that a decrease of the room temperature conductivity from $\simeq 400\Omega^{-1}cm^{-1}$ to $\simeq 320\Omega^{-1}cm^{-1}$ corresponds to an increase of the RRR by a factor of 2-3 (RRR goes from 2.5 to 6). A decrease from $\simeq 320\Omega^{-1}cm^{-1}$ to $\simeq 220\Omega^{-1}cm^{-1}$ multiplies the RRR by a factor of 5 (from 6 to 30). A further decrease from $\simeq220$ to $\simeq 180 \Omega^{-1}cm^{-1}$ (a relative decrease of only 20\%) multiplies again the RRR by more than 5 (from 30 to 175). In other words, there is about the same ``distance'' in the room temperature conductivity between two samples with $RRR=2$ and 6 as between two samples with $RRR=6$ and 100. The RRR dependence on  room temperature resistivity across the MIT (the MIT occurs\cite{DelahayePRB03} for $RRR\simeq 20$) is far from linear but is close to an exponential law (see Figure \ref{FigureB1} of Appendix B). Therefore the small but monotonous increase of the RRR observed as a function of the Pd content in the Figure 8 of Ref. \cite{FisherPMB02} doesn't mean that a RRR as high as 100  cannot be reached  by  further increasing  the Pd concentration by a few at.\% (and with a suitable annealing, see below). Interestingly  in that plot, the RRR increases faster  for the highest Pd contents, in agreement with the approach to a MIT. An increase of $\simeq$ 3 at.\% in Pd has already reduced the zero temperature extrapolated conductivity by a factor of 3, from $\simeq 500\Omega^{-1}cm^{-1}$ to $\simeq 150\Omega^{-1}cm^{-1}$. It is not clear if the highest Pd content reached in Ref. \cite{FisherPMB02} corresponds to the maximum Pd content possible for single-grains or if single-grains with more Pd (and less Al) could actually be made.

As final remarks, note firstly that all the single-grains measured in Ref. \cite{FisherPMB02} have different Al and Pd contents but almost the same Re content. We know that the Re concentration   strongly influences the resistivity of polycrystalline samples (see subsection \ref{CompositionOrder}). Whether a change in the Re content can result in single-grains with larger resistivity and RRR values thus remains an open question.
 Secondly a composition gradient of 0.5 at.\% is mentioned between the centre and the outer part of the single-grains \cite{FisherPMB02}. The central region has a slightly higher Pd content whereas the edges have a slightly higher Re content. It would be interesting to know how the single-grain samples used for the electrical measurements were cut relative to this composition gradient, since composition changes, even minute, certainly affect the measured properties close to the MIT.

\subsubsection{Single-grains versus polycrystalline samples:  effect of annealing}

Annealing of single-grains by keeping them  at $900^oC$ in the melt during few days prior to decanting didn't  lead to  significant change in the RRR or in the structural quality\cite{FisherPMB02}  (but according to the Figure 8 of Ref. \cite{FisherPMB02}, these samples have the largest RRR).  Annealing of  single-grains post-growth outside the melt resulted in partial decomposition of the material \cite{FisherPMB02}. In comparison, post-growth annealing was found to be essential to get high resistivity in all  polycrystalline samples  (see subsection \ref{Annealing}). The optimal annealing parameters depend on the type of samples, i.e. ingots, ribbons or thin films.  Whether  the i-AlPdRe single-grains resistivity can be increased by annealing, like for polycrystalline samples, is still an open question, but the comparison with  i-AlPdMn single-grains is suggestive.

In i-AlPdMn, single-grains with conductivities as low as that found in best quenched and annealed i-AlPdMn ribbons \cite{LancoEPL92} was a many year effort. Not only  chemical composition tuning but also single-grains post-growth annealing  are crucial \cite{PrejeanPRB02}. Depending on the annealing parameters (annealing plateau temperature, cooling rate, etc.) the room temperature conductivity can be reduced from $\simeq 500\Omega^{-1}cm^{-1}$ to $\simeq 300\Omega^{-1}cm^{-1}$ for a given composition. This conductivity decrease was shown to be related to subtle changes of the icosahedral order, the lower the number of local defects, the smaller the conductivity \cite{PrejeanPRB02}. We  come back to this question below (see subsection \ref{CompositionOrder}).

\subsubsection{i-AlPdRe versus i-AlPdMn alloys}

Beyond any sample type consideration (ribbons, ingots, thin films or single-grains), the resistivity values of different i-Al based alloys points to a lower conductivity in AlPdRe than AlPdMn.
In Table \ref{TableAlCuPdTM}, the lowest conductivity  and corresponding highest RRR values are compared for i-Al-Cu-TM and i-Al-Pd-TM alloys (TM is a transition metal with an incomplete d band). The table clearly shows that an increase of the TM atomic number is associated with a decrease of the conductivity and a corresponding increase of the RRR \cite{GrenetAussois}. It has been suggested that heavier TM elements give rise to a stronger scattering and favour electron localization. The effect is more pronounced in the i-Al-Pd-MT alloys but is consistent with what is observed in the i-Al-Cu-MT alloys.

\begin{table*}[htb]
\caption{\label{TableAlCuPdTM}
The lowest 4K conductivities and the highest RRR for selected Al-based ternary icosahedral alloys (reproduced from Ref. \cite{GrenetAussois}).}
\begin{tabular}{|c|c|c|c|c|}
  \hline
  Al-Cu- & Fe & Ru & Os  \\
   & $\sigma_{4K}=100\Omega^{-1}cm^{-1}$ & $\sigma_{4K}=40\Omega^{-1} cm^{-1}$ & $\sigma_{4K}=7\Omega^{-1} cm^{-1}$  \\
   & $RRR = 2.2$ & $RRR = 4$ & $RRR = 4.5$ \\ \hline
  Al-Pd- & Mn &  & Re  \\
   & $\sigma_{4K}=100\Omega^{-1} cm^{-1}$ &  & $\sigma_{4K}=1\Omega^{-1} cm^{-1}$  \\
   & $RRR = 2.3$ & & $RRR > 200$ \\ \hline
\end{tabular}
\end{table*}

\subsection{Argument 2: the role of voids and oxygen rich bridges in ingots}

In Ref. \cite{DolinsekPRB06},  the morphology and  chemical composition of a single-grain and of an ingot with  RRR=154  were investigated
by scanning electron microscope imaging (SEM) and energy dispersive X-ray spectroscopy (EDXS). SEM images of the ingot shows a porous morphology with voids and oxygen rich bridges. The co-occurrence of these two features was then argued to be  the reason for the measured high resistivity. We propose  below an alternative explanation for the  observed oxygen enrichment  in  the ingot bridges, and  a counter argument to an oxide-induced high resistivity in the i-phase.

\subsubsection{The oxygen-rich bridges: an alternative interpretation}

The reported oxygen increase  \cite{DolinsekPRB06} in the narrow bridges of a polycrystalline ingot can be questioned and here we explain  why. The composition of the i-AlPdRe phase was locally investigated using EDXS and according to the authors, a subsurface volume of $0.6\mu m$ diameter was probed. This size is of the same order as the width of the narrow bridges (typically $1\mu m$, see Figure 4 of Ref. \cite{DolinsekPRB06}). The ``background'' oxygen contribution of the polished surface was found to be of 2 at. \% for a single-grain and 3-4 atomic \% for the polycristalline ingot far from the bridges. According to the authors of Ref. \cite{DolinsekPRB06}, this ``background'' contribution represents the oxygen content near the surface. If this is true, we expect an oxygen increase when the bridges are measured due to the additional surface contribution of the bridge edges. Moreover, only the top surface of the sample is polished and not the edges surrounding the bridges on all sides. Due to the high T treatment of the ingot, a thicker oxide build up is thus expected on the edges than on the reference polished surface.
Since in Ref. \cite{DolinsekPRB06} this ``edge'' oxide contribution was not quantified, it is difficult to know  by how much the ``bulk'' of the ingot bridges is effectively enriched in oxygen.

\subsubsection{The resistivity of the oxidized i-AlPdRe phase}

Due to their porous morphology, the ingots can be viewed as a network of large islands connected by narrow bridges. Take the case where bridges and  islands have  different electrical properties.  If  the narrow bridges are much more resistive than the islands, their contribution will dominate the measured resistivity. But the opposite is also true: if the islands are much more resistive than the narrow bridges, the islands will dominate the resistivity. Thus, the observation of high resistivities cannot indicate if the  narrow bridges are more resistive than the islands or vice-versa, or of a different mechanism altogether.

Now assuming, for the sake of the argument, that the bridges are actually more oxidized than the islands. It is often the case that oxidized materials are  more electrically insulating than pure elements and alloys. For example, in granular aluminium, crystalline aluminium grains are separated by amorphous alumina layers. In that case there is a mixing of two different materials (there is no solubility of oxygen in aluminium). An increase of the oxygen content leads to thicker insulating alumina barriers between the metallic grains and thus to an increase of the resistivity \cite{GrenetEPJB07}. But in lack of further evidence, it cannot be absolutely ruled out that oxygen atoms are diluted in the icosahedral structure (although that would be a unique situation). Even in that unlikely case,  the inclusion of oxygen atoms impurities  would most probably \emph{decrease} the i-phase resistivity. Foreign atom in the i-phase structure should act like any other defect that decrease the resistivity, according to the ``inverse Mathiessen rule'' largely demonstrated in Al-based  i-phase, including  i-AlPdRe ingots (see subsection \ref{CompositionOrder}).

\begin{figure}[htb]
\includegraphics[width=7cm]{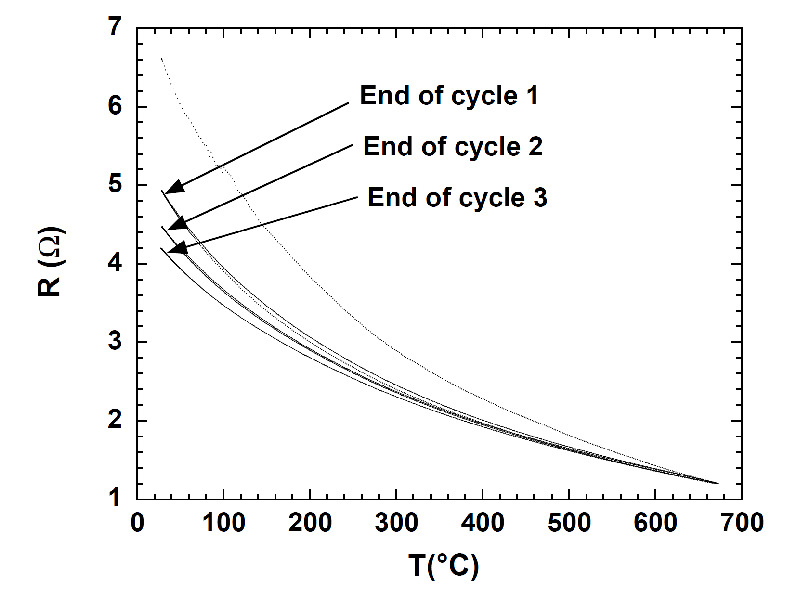}
\caption{\label{FigureCECMCyclesT} Resistance versus temperature for an i-AlPdRe melt-spun ribbon during three temperature cycles in air between 25 and $700^oC$ (temperature ramps of $100^oC/h$), starting from a resistance of $R_{300K}=6.6\Omega$ to end at $4.2\Omega$}.
\end{figure}

There are a number of experimental results which indeed suggest that an oxidation of an i-AlPdRe sample reduces its electrical resistivity. In their comment \cite{PoonPRB07}  to Ref. \cite{DolinsekPRB06}, Poon and Rapp mentioned that a reduction of the base pressure during the fabrication of i-AlPdRe ingots leads to a strong increase in the resistivity and the RRR values (see also Ref. \cite{PoonMRS99}). In an i-AlPdRe ribbon, we have observed that the RRR is gradually reduced from 100 to 6 after three successive heat treatments in air up to $700^oC$ (see Ref. \cite{DelahayeDEA97} and Figure \ref{FigureCECMCyclesT}).

To demonstrate that the high resistivity values of the polycrystalline ingots are due to a high oxygen content of their narrow bridges would require to show a systematic morphology or oxygen content correlation with the RRR (only one ingot was studied in Ref. \cite{DolinsekPRB06}). Furthermore one also has to understand the case of ribbons and thin films.
Extensive SEM imaging (see Appendix \ref{AppendixA}) shows that  ribbons have no porosity. But because ribbons are  only a few tens of micrometres thick, we have to considered if  oxygen could diffuse to the bulk of the ribbon  during  melt-spinning  and  subsequent   high temperature annealing. Firstly deep oxidation is unlikely because ribbons are annealed  in sealed quartz tubes that are first pumped  down to high vacuum. Secondly we have observed ribbons melt-spun from the same melt, annealed together in the same tube and at the same temperature with RRR as different as 2 and 200 (see subsection \ref{Annealing} and Appendix \ref{AppendixA}).  It is hardly conceivable that  samples processed  in the exact same batch would contain the very large difference in their oxygen content required  to solely explain their change in RRR.

The oxidation state of the i-AlPdRe samples and its influence on the resistivity values is  nonetheless an important question that has been rarely considered and  deserves further studies. Nothing is known about the oxygen atoms position in the structure. Are they diluted in substitution to the i-phase? Are they concentrated in the grain boundaries? The answers to these questions are interesting but far beyond the resolution of previous SEM investigations.

\subsection{Argument 3: the DOS of AlPdRe compared to AlPdMn}

In Ref. \cite{DolinsekPRB06} Dolin\v sek et al compared the NMR measurements  on a single-grain of $RRR \simeq 1.2$ and on an ingot of $RRR = 154$. The local DOS at the Al sites are about the same in both samples, which cannot explain the large difference observed in the resistivity values. They also performed  ab-initio calculations on isomorphous i-AlPdMn and i-AlPdRe alloys, which show that the DOS around the Fermi level  ($E_F$) of the two compounds are very similar. They concluded  that similar  DOS cannot explain the large difference observed in the resistivity values that should therefore be attributed to an extrinsic origin.  We show below why this conclusion is questionable by discussing first other DOS measurements  in i-AlPdRe that we then put into context of the  MIT in disordered insulators.

Firstly, the DOS of the icosahedral alloys has been measured by several  techniques (NMR, specific heat, photoemission and  tunneling spectroscopy) and the experimental situation is more complex, and even sometimes controversial,  than a single NMR measurement on only two samples may  indicate (the same is also true for  theoretical studies of the DOS). In NMR measurements, no significant difference is usually found between highly resistive i-AlPdRe, i-AlCuFe or i-AlPdMn samples \cite{SimonetICQ6, TangPRL97}. But difficulties in the interpretation of the results were  raised \cite{SimonetICQ6}. According to specific heat measurements, the DOS around $E_F$ is lower in i-AlPdRe than in the other quasicrystals. The DOS is one tenth that of pure aluminium for i-AlPdRe  compared to one third for i-AlCuFe and i-AlPdMn samples \cite{PiercePRL94, ChernikovEL96, PrejeanPRB00}. The absence of a linear term below 1K in a ribbon of RRR = 80 was ascribed to  the localized nature of the electronic states \cite{PrejeanPRB00}. In tunneling spectroscopy measurements, the zero bias anomaly observed in the $dI/dV$ curves is  more pronounced in the i-AlPdRe samples than in the other i-phases \cite{DavydovPRL96, DelahayePRB03}. In low temperature photoemission measurements, the comparison between the different samples is more difficult due to the strong influence of the surface preparation \cite{SchaubEPJB00}. But for samples prepared similarly, a significantly lower spectral intensity at $E_F$ is observed for i-AlPdRe compared to i-AlCuFe and i-AlPdMn \cite{StadnikPRL96}. Moreover, several authors have underlined  problems in interpreting  the specific heat and tunneling spectroscopy measurements as usual DOS when the system is close to the metal-insulator transition \cite{PrejeanPRB00, DelahayePRB03}.

Secondly, we emphasize that small change in the DOS at $E_F$ does not preclude large increase in the low temperature resistivity. According to the Einstein's formula, the conductivity is proportional to the DOS at $E_F$, but also to the electron diffusivity. In the theoretical framework of the Anderson MIT, the DOS at $E_F$ remains finite in the insulating state as the electronic wave functions become localized. It is thus the electron mobility and not the DOS that vanishes at zero temperature. The  non semi-conducting behavior (no band gap) of i-AlPdRe samples and the strong similarities between the electronic properties of i-AlPdRe samples and disorder systems close to the MIT clearly indicates that the electron diffusivity plays the crucial role.

Like for the DOS measurements, the interpretation of the thermoelectric power must be taken with great caution. In Ref. \cite{DolinsekPRB06}, the authors have found that an insulating i-AlPdRe polygrain has a large and positive Seebeck coefficient while this coefficient is small and negative in a metallic single-grain. But large and positive Seebeck coefficients are also found in some metallic i-AlCuFe and i-AlPdMn samples
\cite{Giroud96,PiercePRB93}, and thus cannot be used to differentiate the insulating i-AlPdRe polycrystalline samples from all the other metallic i-Al based alloys.

\subsection{Argument 4: the role of secondary phases}

A minority of secondary phases are indeed found in ingots and ribbons \cite{DelahayeJNCS99,RodmarJNCS99,RodmarPRB00}(the existence of such secondary phases is not mentioned in thin films). But  contrary to the claim in Ref. \cite{FisherPMB02}, these secondary phases cannot explain the high resistivity values observed. Firstly, from the SEM images they don't percolate across the samples and  have a small volume fraction. Reminding that the  ribbon samples are non porous (see discussion above), a small fraction of secondary phase  cannot induce an insulating  character of the samples.  Secondly, these secondary phases are alloys of metals (composition close to $Al_{73}Re_{27}$ in the bulk of the ingots \cite{RodmarJNCS99,RodmarPRB00} and to $Al_{50}Re_{50}$ at the surface of the ribbons \cite{DelahayeJNCS99,RodmarPRB00}). They are most likely less resistive than the i-phase itself (the opposite would be noteworthy!). Thirdly, no correlation between the amount of  secondary phases and the sample RRR could be found.

\section{\label{PolycrystallineSamples} Origin of the high resistivity values: the polycrystalline sample results}

Following the arguments in the previous section, the low resistivity of single-grains  doesn't imply an extrinsic origin of the high resistivity of  polycrystalline i-AlPdRe. In this section, we propose a short review of the large set of results  obtained by different groups on polycrystalline samples.  We show that there are indeed a number of experiments in favour of an intrinsic origin. Even if the parameter(s) that control the MIT have yet not  been clearly identified, the chemical composition and the structural order quality of the i-AlPdRe sample seem to play a determining role in the high resistivity.

\subsection{\label{Annealing} The essential role of annealing}

 According to the literature, high resistivity values can only be obtained after a specific annealing process. Since  annealing is used to homogenize the chemical composition and to improve the structural order, it is already a strong indication that a well defined chemical composition and a high structural quality are associated with the high resistivity values.

After melt-spinning, ribbons are not homogeneous and contain a large amount of secondary phases, as highlighted in the SEM picture of Figure \ref{FigureA1} in Appendix \ref{AppendixA}. These ribbons have room temperature conductivities in the range $1000 - 3000 \Omega^{-1}cm^{-1}$, i.e. about ten times larger than after annealing, and RRR around 1. A heat treatment is  necessary for the atoms to interdiffuse and homogeneize and to obtain quasicrystalline grains. Details are given in Appendix A. There isn't a single way to get RRR values larger than 100. The ribbons can either be annealed for a long time ($\simeq 6h$) at a not too high T ($\simeq 900^oC$) (``low T'' ribbons) or be annealed for a much shorter time ($\simeq 10mn$) but closer to their melting point (up to $1020^oC$) (``high T'' ribbons). The latter case includes slow temperature ramps ($50-100^oC/h$). The highest temperature reached during the heat treatment clearly influences the RRR distribution among samples annealed in the same batch (see Figure \ref{FigureA3} of Appendix \ref{AppendixA}).  It is noteworthy  that ``high T'' ribbons that are melt-spun from the same ingot and annealed at the same time can have RRR ranging from 2 to 200.

Before annealing, as-cast ingots are mainly composed of crystalline phases with compositions relatively far from the icosahedral domain \cite{GignouxThesis96, KiriharaMRS99} and a heat treatment is  necessary.
There is an agreement in the literature on the heat treatment to get high resistivity and large RRR values. An annealing at high temperature (between $860^oC$ and $980^oC$) during about ten hours gives X-ray diffraction patterns that can be indexed by an icosahedral phase \cite{TsaiMTJIM90, PierceScience93, LinJPCM97} and maximum RRR around 30 \cite{PierceScience93}.
A full chemical homogeneity of the ingots is however not achieved (see subsection \ref{CompositionOrder}) and  porosity  can reach $40\%$. An additional annealing to lower temperature (a few hours between $600^oC$ and $750^oC$) is needed to increase the resistivity values (the 4K resistivity is multiplied by about 5) and RRR above 100 (the record being 280) \cite{PiercePRL94}. This low temperature transformation was found to be reversible since a further annealing of the ingots above $750^oC$ restores RRR around 20 \cite{PiercePRL94}. We will come back to the possible interpretation of these findings at the end of subsection \ref{CompositionOrder}.  Nothing similar was observed in our ribbons: a further annealing step at $600^oC$ leads to a decrease of the RRR (see Table \ref{TableRecuitCECM} of Appendix \ref{AppendixA}).

For thin films, the annealing parameters are very different. The films obtained by the co-evaporation of Al-Pd and Re are amorphous before annealing \cite{HaberkernCondmat99}. After a few hours at $730^oC$, the films are composed essentially of the icosahedral phase  with a large distribution of grains sizes from 40nm to $1\mu m$. Haberkern et al have observed an increase of the 4K resistivity by a factor of 30 after annealing and an increase of the RRR up to 20 \cite{HaberkernCondmat99}. A similar heat treatment is reported in the films made by the sequential evaporation of Al, Pd and Re, with a maximum RRR of only 4 \cite{BergmanStage99}.

\subsection{\label{CompositionOrder} Chemical composition versus structural order}

The strong sensitivity of the electrical resistivity to chemical composition and presence of  defects has been studied in details in i-AlCuFe and i-AlPdMn alloys. In both alloys, the highest resistivity values (and the highest RRR) are observed for a well defined chemical composition. Any departure from this composition, even by only 1 at.\%, leads to a lower  resistivity. For example, in i-AlCuFe, a change in the Fe concentration by only 0.5 at.\% doubles the 4K conductivity \cite{LindqvistPRB93}. Moreover, for a given chemical composition, the smaller the defect density, the larger the resistivity \cite{MayouPRL93}.
This so-called ``inverse Mathiessen rule'' was clearly demonstrated in i-AlPdMn single-grains where  defects are associated with the apparition of Mn magnetic moments and can be probed by magnetic measurements down to very low concentrations \cite{PrejeanPRB02}.
Subtle changes in the structural order of the i-AlPdMn phase (F, F2 and F2M structures) were also found to be at the origin of significant resistivity changes \cite{PrejeanPRB02}.
We  show below   clear indications that the resistivity in i-AlPdRe samples is also very sensitive to chemical composition changes and the presence of defects.

\subsubsection{\label{ChemicalComposition}  Role of chemical composition}

The most convincing evidence for a direct relation between the chemical composition of the i-AlPdRe phase and its resistivity was obtained in thin films made by the co-evaporation of Al-Pd and Re \cite{HaberkernCondmat99, HaberkernICQ7}. With this technique, the chemical composition of the thin films can be systematically varied  (the absolute composition is not precisely known since the composition is only estimated from the relative thickness of the elemental components). In  Figure 6.1 of Ref. \cite{HaberkernCondmat99}, the low temperature conductivity value (1.3K) is the lowest at a Re content of 7.5 at.\%. A deviation of the Re content from this value by only 0.5 at.\% increases the low temperature conductivity by a factor of 4. The room temperature conductivity shows a monotonous increase with the Re content.

In ribbons and ingots, there is an agreement in the literature  that high resistivity and RRR values are obtained only for a given nominal composition. The reported ``optimal'' nominal composition has sometimes changed over time and is not universal (see Table \ref{TableSamples} and Ref. \cite{RappBook99}). Moreover, large sample to sample  resistivity and RRR variations  are often observed for a given nominal composition. This  apparent contradiction can be resolved noting that reported nominal compositions  may differ significantly from the actual ones and  the samples are not perfectly homogeneous. Due to the high melting point of Re ($3186^oC$) compared to that  of Al ($660^oC$) and Pd ($1558^oC$), the fabrication of homogeneous i-AlPdRe ingots and ribbons with a controlled chemical composition is more difficult than for i-AlCuFe and i-AlPdMn alloys.

In i-AlPdRe ingots, Sawada et al \cite{SawadaICQ6} have explored a large composition range within and around the single i-phase domain. The higher resistivity and RRR values are found in the single i-phase region and within this region, for a ``e/a''  ratio (electron per atom ratio \cite{TsaiBook99}) approaching 1.78 and when the Al concentration decreases. For some samples,  the actual chemical composition was measured and presents systematic shifts of the order of 1 at.\% relative to the nominal ones. The shifts  are not always reproducible \cite{SawadaICQ6}.

We have performed local chemical investigations of i-AlPdRe ribbons and ingots having different RRR. Our goal was to evaluate quantitatively the homogeneity of the samples and to possibly identify a common composition range for  samples having high RRR. We have used a Castaing microprobe which probes a volume of only a few $\mu m^3$. This  provides  measurement of the chemical composition at the scale of individual grains. The accuracy on the chemical composition (systematic and random errors) is of the order of 1 at.\%, and the precision (random error) is a fraction of 1 at.\%. We have performed chemical composition profiles in the length,  width and  thickness of polished samples, both ribbons and ingots.

\begin{figure}[h]
\includegraphics[width=8cm]{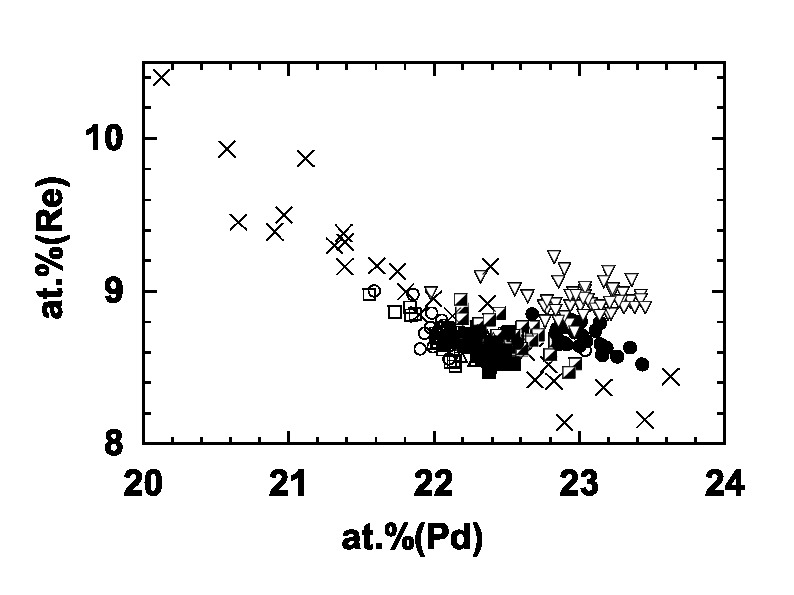}
\includegraphics[width=8cm]{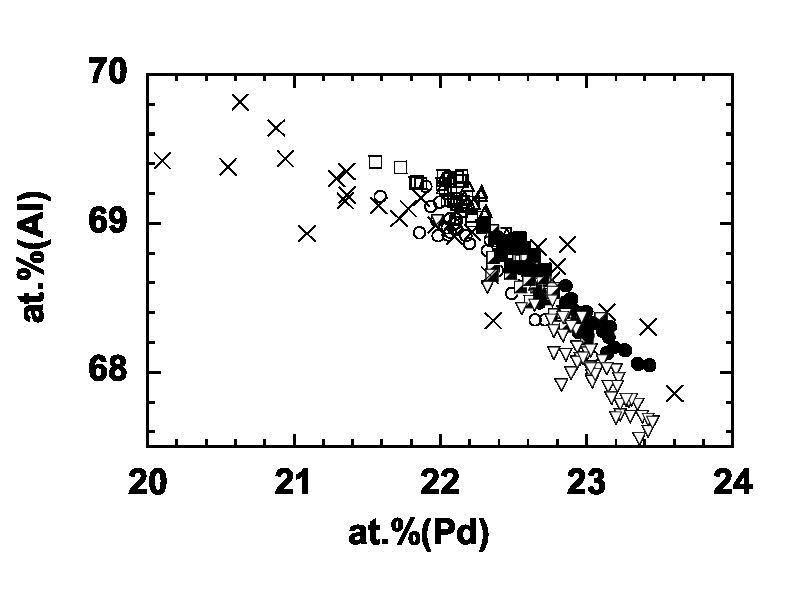}
\caption{\label{Figure2} Chemical composition domains explored by six i-AlPdRe ribbons of nominal composition $Al_{70.5}Pd_{21}Re_{8.5}$. Cross symbols: ``low T'' ribbon (RRR = 100). Other symbols: ``high T'' ribbons ($T_{max}=1020^oC$). Empty symbols: $RRR > 90$; half-filled symbols: $RRR = 25$; filled symbols: $RRR < 5$. The accuracy is about 1 at.\% for the different chemical elements. The precision is  0.5 at.\% for  Al, 0.2 at.\% for  Pd and 0.15 at.\% for  Re.}
\end{figure}

The length profile results (step size between $25$ and $100\mu m$) obtained on five ``high T'' i-AlPdRe ribbons melt-spun from the same ingot are plotted in Figure \ref{Figure2} (all but cross symbols). Conclusion are as follows. First, the actual  composition is significantly different from the nominal one, which was $Al_{70.5}Pd_{21}Re_{8.5}$ for all the ribbons. The Al content is lower (67.5-69.5 at.\%) and the Pd and Re contents higher than expected (21.5-23.5 at.\% for  Pd, 8.5-9.2 at.\% for  Re). Second, the Pd and Al concentrations are strongly correlated, while no clear correlation appears between the Re and the Pd contents. Composition variations can occur either on the length scale of a millimetre or on a much smaller scale (of the order of the grains size). A  composition gradient is often present in the thickness of the ribbons with typical variation of 1 at.\% for  Al and  Pd but no significant change for  Re (data not shown). Third, high RRR ribbons cannot be associated with a single  composition.
Forth, the high RRR values are not necessarily observed in the samples which are the most homogeneous. Low RRR ribbons  often have smaller composition fluctuations  than the high RRR ribbons. These conclusions are solid and have been confirmed (not shown) by  the microprobe analysis of many more ``high T'' ribbons that were melt-spun from different ingots.

The role of the maximum annealing temperature on the  chemical homogeneity of ribbons is also presented in Figure \ref{Figure2}. The composition domain spanned by a ``low T'' ribbon annealed 6h at $900^oC$ and of RRR = 100 (cross symbols) is compared to ``high T'' ribbons (other symbols). The ``high T'' ribbons have much smaller Re and Pd content fluctuations than the ``low T'' ones, whatever their RRR values.
The slopes of the Al - Pd and the Re - Pd correlations seem also to be different between the two types of ribbons. The results of Figure \ref{Figure2} demonstrate that high RRR can be found in quite inhomogeneous ribbons.

\begin{figure}[h]
\includegraphics[width=8cm]{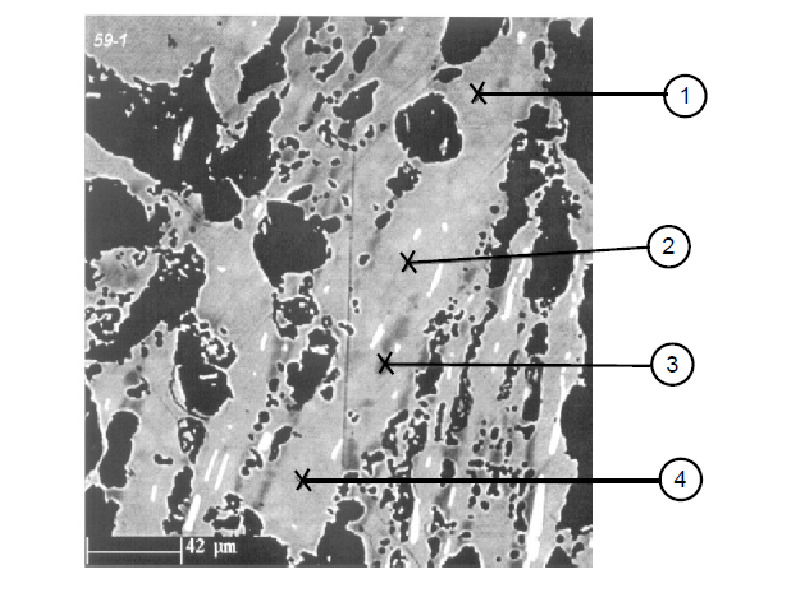}
\caption{\label{Figure3} SEM picture (backscattered electron mode) of a polished ingot of RRR = 59. The contrast visible in the main phase (in grey) corresponds to chemical composition fluctuations of a few at.\%. The microprobe analysis at the location of the x's  in the picture give for Al, Pd and Re in at.\%: point 1: 68.4, 22, 9.7; point 2: 68.7, 22.8, 8.6; point 3: 66.9, 25.9, 7.2; point 4: 68.1, 23.6, 8.3. The  precision of the compositions is estimated to be a few 0.1at.\%.}
\end{figure}

\begin{figure}[h]
\includegraphics[width=8cm]{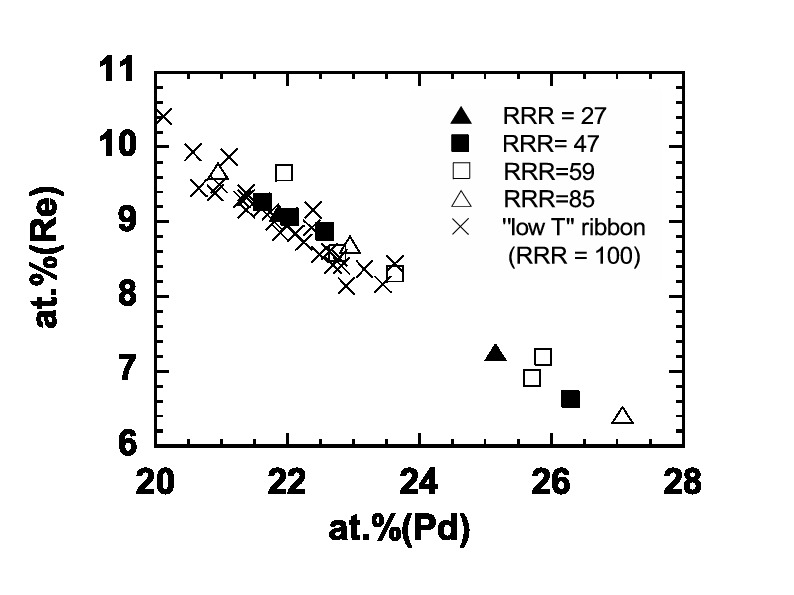}
\includegraphics[width=8cm]{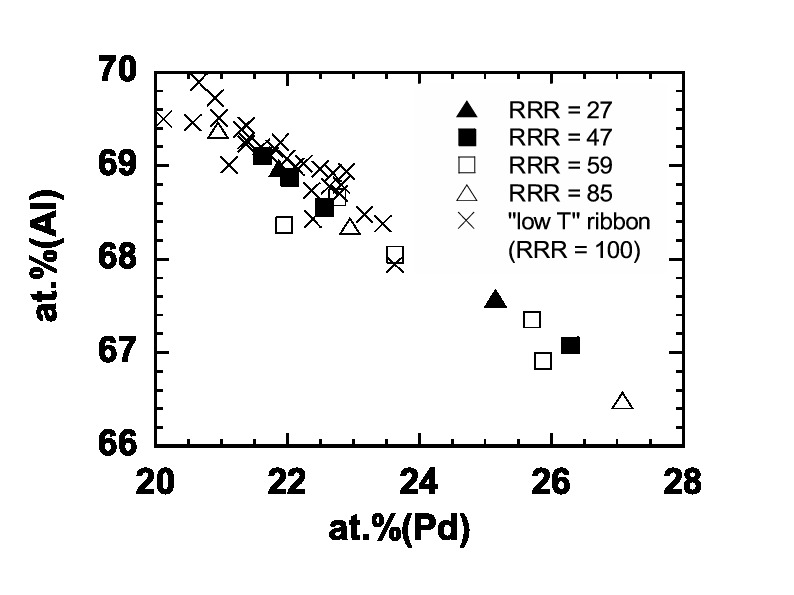}
\caption{\label{Figure4} Chemical composition range of  4 ingots with RRR between 27 and 85. The ``low T'' ribbon of Figure \ref{Figure4} annealed at $900^oC$ (RRR = 100) is also plotted for comparison (x symbols).}
\end{figure}

We have also measured the  composition fluctuations on 4 ingots, with RRR between 27 and 85. The composition variation of the main phase is quite large and ranges between 67 to 69 Al at. \% , 21-27 Pd at.\%  and 7-10 Re at.\% (see the SEM picture of Figure \ref{Figure3}). We did not find any significant difference either in the amplitude of the fluctuations or in the composition range between  ingots of different RRR (see Figure \ref{Figure4}). Interestingly, the  composition domain of ingots is similar to that of the ``low T'' ribbon of Figure \ref{Figure2}. This   indicates that composition of the i-phase has to do with  the maximum annealing temperature more than the sample type, i.e. ingots or ribbons (all the ingots of Figure \ref{Figure3} and \ref{Figure4} have been annealed to a maximum temperature of $\simeq 900^oC$).
The  composition domain  in Figures \ref{Figure2} and \ref{Figure4} coincides roughly with that of the single i-phase  in the ternary phase diagram \cite{SawadaICQ6} which explain why the X-ray diffraction peaks can be indexed by the i-phase, despite composition variations.

The fact that high resistivity values are found in inhomogeneous ribbons and ingots  seems contradicting  the existence of a well defined optimal composition for the highly resistive i-phase. However the effect of the composition fluctuations on the resistivity have to be considered together with  the microstructure of the samples; more precisely the typical length scale of the composition  fluctuations should be compared with the microstructure. In a porous sample, even a small volumic fraction of highly resistive areas  can dominate the sample resistivity if their extend is larger than the ``bridge'' length scale (see argument above). Due to the microstructure of the sample, the current will pass through these highly resistive zone. In ingots, important chemical composition fluctuations are observed within the dendrites which may include the optimal composition associated with the high resistivity values. The results on the ribbons which have no porosity are  more difficult to reconcile with the hypothesis of an optimal chemical composition. If any departure from this optimal composition gives rise to a strong decrease of the resistivity, inhomogeneous samples like the ``low T'' ribbon of Figure \ref{Figure2} can only be highly resistive if the low resistivity regions do not percolate. We will give indication for electrical inhomogeneities in the ribbons in subsection \ref{NonUniversal}.

\subsubsection{The role of defects}


Like in any other crystalline material,  defects in a quasicrystalline sample can be chemical (impurities, substitution of an atomic species by another, etc.) or structural (atomic position displacement, vacancies). As we will shortly review below, many experiments performed on i-AlPdRe polycrystalline samples suggest that, like for the other quasicrystalline alloys, the lower the defect density, the larger the resistivity.

Following subsection \ref{Annealing},  annealing of ingots and ribbons modifies not only the structural order  but also  the  composition homogeneity. It is therefore difficult in these samples to distinguish  the respective effect of  structural improvment and of composition homogeneization. In i-AlPdRe thin films made by the co-evaporation of Al-Pd and Re, the situation is different. Co-evaporated  films are amorphous and   undergo a transition to a quasicrystalline film upon anneling. This  can be  followed as a function of  annealing time \cite{HaberkernCondmat99, HaberkernICQ7}. Like in the i-AlCuFe and i-AlPdMn alloys, the resistivity  increases when the  icosahedral order quality is improved. The resistivity of the amorphous phase at room temperature is about one order of magnitude lower than that of the icosahedral phase, and a clear  inverse Mathiessen rule is observed in the conductivity.

The role of  structural defects in reducing the resistivity and RRR of i-AlPdRe phases was also revealed  by neutron irradiation studies of ribbons and ingots \cite{KarkinPRB02,RappJPCM08}. The resistivity of AlPdRe samples was found to decrease strongly with neutron irradiation: the larger the dose, the lower the resistivity. The RRR can decrease from a value of 100 before irradiation down to $\simeq 1$ in irradiated samples. Analysis of  X-ray diffraction peaks indicates  that neutron irradiation preserves the icosahedral symmetry but introduces structural defects  in the icosahedral phase and decreases the coherence length of the icosahedral order. Relatively low temperature annealing of the irradiated samples  (below $600^oC$)  shouldn't  affect the overall composition of the sample yet it  increases the resistivity and RRR (up to $\simeq 2$) \cite{RappJPCM08}.

More subtle effects may be important. Beeli et al \cite{BeeliICQ7} have compared the structural order of  i-AlPdRe ingots by transmission electron microscopy (TEM). The samples had two different heat treatments: annealed only at high T (``HT'' ingots), with RRR below 30, or annealed at high and subsequently  low T (``LT'' ingots) with RRR above 100. While the ``HT'' ingots are close to a perfect i-state with only some random phason strain, a strongly phason-disturbed i-phase was observed in the ``LT'' ingots. These results have not been confirmed by any other group since. It would be interesting to perform similar investigations  to know if  high RRR ribbons have also a phason-perturbed i-phase. By comparison, the resistivity of AlPdMn single-grains was found to be sensitive to the existence of icosahedral superstructures (F2, F2M). Whether similar superstructures play also a role in the resistivity values of i-AlPdRe samples remains an open question.

The role of  defects was more specifically studied for chemical substitution in i-AlPdRe samples, with Re substituted by  Mn \cite{GuoPRB96,LinJPCM96} or Ru \cite{WangSSC98}. All reports agree on a decrease of both resistivity and  RRR with an increase of the Mn or Ru content. The resistivity change is less pronounced for Ru than for Mn substitution and high RRR values could still be observed for high Ru contents (a RRR of 38 was observed in Ref. \cite{WangSSC98} for a 40\% Ru substitution). Unfortunately in this case the disorder was not well characterized and it is not clear if the substitution results in a mixture of different phases or in a ``true'' microscopic chemical disorder.

The presence of a small amount of chemical impurities  also influences the maximum  RRR values. Poon et al \cite{PoonMRS99} indicate that the maximum RRR depends on the ``cleanness''  of the environment during the melting of the metallic elements and during  annealing. After  cleanness improvement (not precisely quantified), the maximal RRR are in the range 170-280 compared to 100-170 before. A similar trend was also observed in our ribbons. In a set of ribbons containing 0.3 at.\%Si, 0.25 at.\%Cu and 0.1 at.\%Fe, the maximum RRR was about 130, compared to more than 200 in an other set of ribbons having only 0.1 at.\%Si and no detectable Cu and Fe (below 0.05 at.\%Fe and below 0.06 at.\% Cu -these values were measured with the Castaing microprobe). The statistical significance of these results  has to be taken with caution though, since we have measured the impurity levels only in a small number of ribbons. Note that most reports on AlPdRe samples  give only the purity of the starting metallic elements (Al, Pd, Re) as a cleanness criterion, and do not account for possible impurities  introduced during the fabrication and  annealing process.

\subsection{\label{NonUniversal} Non universal behavior of highly resistive i-AlPdRe samples}

We reported in the previous subsection that the morphology and  chemical homogeneity  are very different for ribbons and ingots: porosity and large chemical composition fluctuations are observed in ingots, whereas no porosity and smaller chemical composition fluctuations with gradients in the thickness for the ``high T'' ribbons. Therefore  the $\sigma(T)$ curves of ingots and ribbons shouldn't  be exactly the same, unless the chemical composition plays no role at all in the   resistivity  (which would be in contradiction with many experimental findings, see subsection \ref{CompositionOrder}). We show below that $\sigma(T)$ indeed differs slightly for ingots and ribbons at low temperature. In particular, although  $\sigma(T)$ and  RRR correlate relatively well \cite{PoonPRB07}, we do observe some difference in $\sigma(T)$ between ribbons and ingots even for the same RRR. This difference, sometimes already visible at room T, is amplified at low T or for samples of large  RRR.

\begin{figure}[h]
\includegraphics[width=8cm]{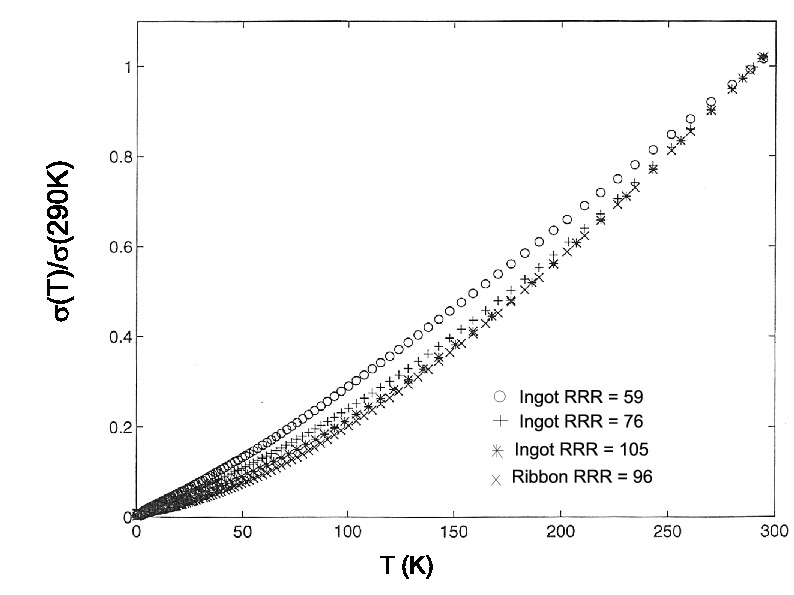}
\includegraphics[width=8cm]{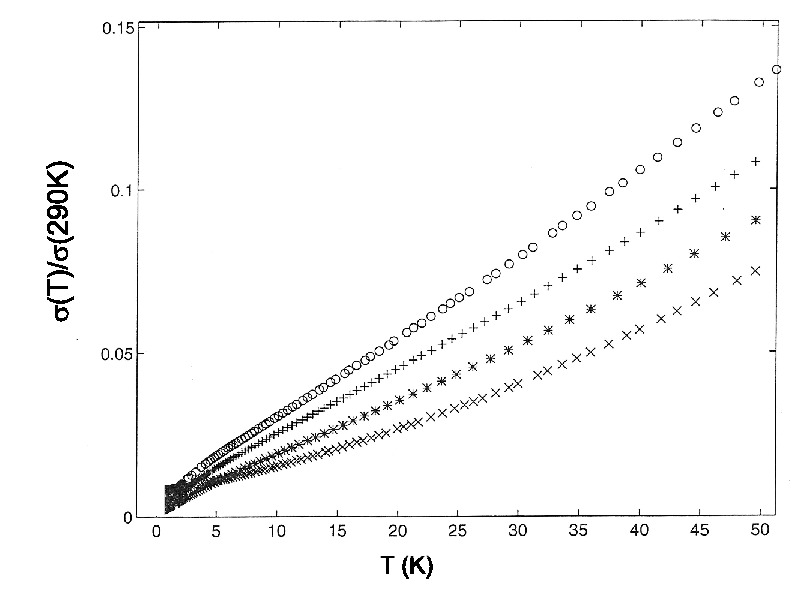}
\includegraphics[width=8cm]{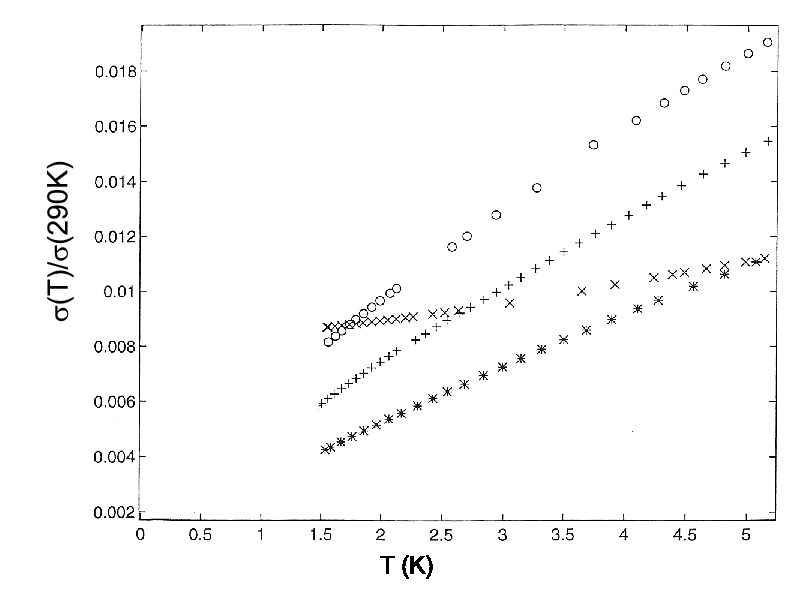}
\caption{\label{Figure5} Normalized $\sigma (T)$ curves for three ingots (RRR = 59, 76 and 105) and one ribbon (RRR = 96) in different T ranges. These curves have been measured in Stockholm (KTH) by O. Rapp and M. Rodmar.}
\end{figure}

A typical difference between ingots and ribbons of high RRR is shown in Figure \ref{Figure5}, where the $\sigma(T)$ curves of three ingots with RRR between 59 and 105 and one ribbon (``high T'') with a RRR at 96 are compared in the $1.5K-300K$ range at  different T scales. At large T scale ($0K-300K$), the overall $\sigma(T)$ curves are rather similar between ribbon and  ingots of similar RRR. Notable differences are observed however by zooming in  ($0K-50K$ or $0K-5K$): the temperature dependence of the conductivity is much more pronounced below $\simeq 10K$ in the ingots than in the ribbon. Moreover, there is an inflexion point for all the ingots in the range $5K-10K$ that is  not present for the ribbon in this temperature range (see also Ref. \cite{DelahayeJPCM03}). This inflexion point is the onset of a stronger conductivity decrease at low T and was  observed in ribbons of high RRR, but only below 1K \cite{DelahayePRL98}. We specifically do not  discuss here the mK temperature range where discrepancies in $\sigma(T)$ dependences are presumably related to questions of  cooling \cite{DelahayeThesis99}. The differences shown in Figure \ref{Figure5} are also found  in  Figure 11 of Ref. \cite{RodmarPRB00} that compares  $\sigma(T)$  for a larger number of ribbons and ingots with RRR between 62 and 119. Interestingly, a careful study of   ribbons (see Appendix \ref{AppendixB}) shows  similar $\sigma(T)$ differences  between ``high T'' (more homogeneous) and ``low T'' (less homogeneous) ribbons. This  strongly suggests that the $\sigma(T)$ differences observed between ingots and ribbons indeed come from differences in their chemical homogeneity (and morphology). The chemical composition gradient observed in the thickness of the ribbons may favour somewhat  parallel conduction of layers  with different conductivity behaviors, whereas the ingots morphology may favour a  conduction in series of small size domains with different resistivity values. This picture is in qualitative agreement with the fact that the $\sigma(T)$  variations are less pronounced at low T in the ribbons than in the ingots of high RRR. Similar  non-universality of the electronic properties have also been observed in some disordered systems close to the metal-insulator transition \cite{MobiusPRB99}.

The reported $\sigma(T)$ similarity  between ribbons and ingots of the same RRR \cite{RappPRB11} may therefore  apply only for samples of moderate RRR value (below 60 in Ref. \cite{RappPRB11}), i.e. in a RRR range where the differences between ingots and ribbons are less pronounced \cite{RodmarPRB00}. In this respect, it would  be interesting to compare  $\sigma(T)$ for the  ingot of RRR = 250 \cite{RappPRB11} with a ribbon of the same RRR. Similarly to our observations, in Ref.  \cite{RappPRB11} an inflexion point is found in the ingots of high RRR around 10K and not in the ribbons.

In thin films, the RRR doesn't exceed $\simeq 20$. In this case also there is indication that the electronic properties are not solely determined by  the RRR  value. In  Figure 6 of Ref. \cite{HaberkernCondmat99} the room temperature conductivity  increases regularly as a function of the Re content, whereas the 1K conductivity goes through a minimum for a Re content of 7.5 at.\%. This alone indicates there is  no one to one correspondence between the RRR and  $\sigma(T)$. Even more interesting are the measurements of the Hall coefficient and the Hall mobility of thin films as a function of their Al content \cite{HaberkernICQ7}. When the Al content  increases from 69 to 76 at.\%, the RRR goes through a maximum for  72.5 at.\% Al whereas the Hall coefficient changes its sign precisely at this composition. Clearly,  samples on both sides of the maximum RRR can have the same RRR and different Hall coefficient signs.


The models developed for homogeneous disordered systems, like Mott's variable-range-hopping conduction theory, have been extensively used in i-AlPdRe insulating samples. But what is the validity and the physical meaning of the extracted parameters if the samples are heterogeneous? Note that in ribbons and ingots of high RRR, the analysis of  $\sigma(T)$  by variable-range-hopping laws gives ``anomalous'' parameters. In ribbons for instance, the Mott's law is observed \cite{DelahayePRL98}, but in a temperature range where the variable range hopping theory should not apply (Mott's temperature below the measurement temperature range). In ingots,  $\sigma(T)$  can only be fitted by variable-range-hopping law if  a constant is added, which physical meaning is unclear  \cite{GuoPRB96,RodmarPRB02,RosenbaumJPCM04}. Deviations from electrical homogeneity in the i-phase may play a role in these anomalies.

\subsection{Highly resistive i-AlPdRe samples as granular metals?}

Even if the i-AlPdRe samples present some electrical inhomogeneity, they cannot be treated as a granular metal with nanometer size islands,  contrary to the suggestion of  Ref. \cite{VekilovEPL09}. In a typical granular metal, metallic grains of nanometric size are usually separated by thin ($1nm$ or so) layers of insulating material. A typical TEM picture of a granular Al film is shown in  Figure \ref{Figure6} below. In this case, crystalline Al grains are embedded in an amorphous insulating alumina matrix which occupies a large volume fraction. No such amorphous phase is observed in structural studies of the diffraction spectra of highly insulating i-AlPdRe samples, either in diffraction or TEM images.
The chemical heterogeneity revealed by SEM or microprobe analysis in Ref. \cite{DolinsekPRB06} occurs at a much larger scale (the quasicrystalline single grains have a typical size of a few micrometers). Moreover,  it would be quite extraordinary that such  electrical heterogeneity would extend from a good metal (like Al) to a good insulator (like alumina), and again no experimental evidence points in that direction. In any case, the actual role of  grains boundaries in the observed resistivity values and $\sigma(T)$  still has to be clarified.


\begin{figure}[h]
\includegraphics[width=8cm]{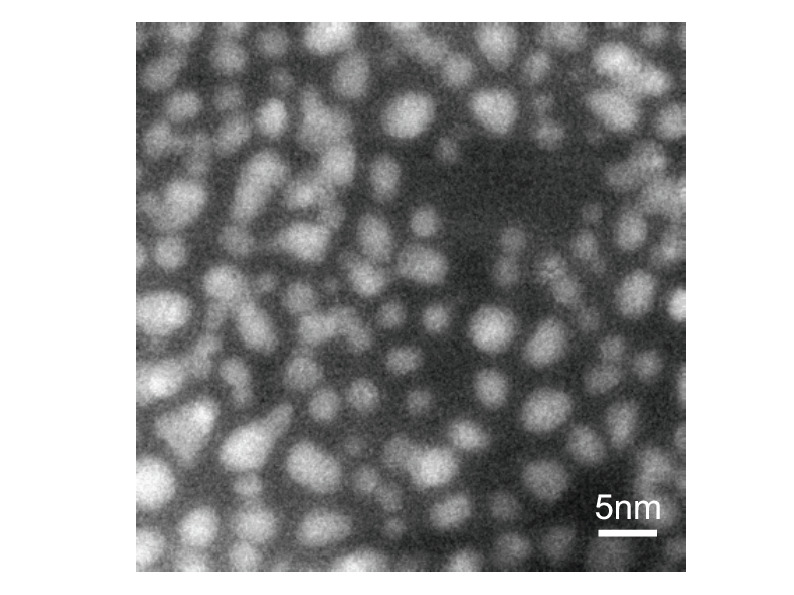}
\caption{\label{Figure6} TEM picture of a granular Al film, 10nm thick and close to the metal-insulator transition. The electrons that have excited the Al plasmons have been selected: the Al grains thus appear in white and the alumina matrix in black. Courtesy of M. C. Cheynet (SIMAP, Grenoble, France).}
\end{figure}

\section{\label{RRR} Conclusion}

In conclusion, a comprehensive study of the experimental results obtained on i-AlPdRe samples (ribbons, thin films,  ingots and single crystals) for this work and from the literature indicates that the high resistivity values  observed in  polycrystalline samples are intrinsic to the i-phase. In particular, the  high resistivity ratio RRR = $\rho_{4K}/\rho_{300K}$ in polycrystalline samples cannot  be explained by the presence of oxide or other secondary phases  \cite{FisherPMB02,DolinsekPRB06,DolinsekPRB07}. In agreement with results in  other quasicrystalline alloys like i-AlPdMn and i-AlCuFe, the resistivity of  i-AlPdRe samples appears to depend on small composition change in the icosahedral phase and on the presence of structural or chemical defects. Specificities of  i-AlPdRe samples may also complicate the picture. Firstly,  the chemical composition of the samples is not well controlled and the samples are never perfectly homogeneous for  metallurgical reasons. Secondly, the samples are on both sides of  a MIT, which amplifies the  sensitivity of transport properties to  any MIT-driving parameter, especially at low temperature. This probably explains why RRR values from 2 to 200 can be observed in batches of i-AlPdRe polycrystalline samples of the same nominal composition and subjected to the same heat treatment. We confirm previous findings that  RRR correlates  relatively well with the conductivity value and its overall temperature dependence, but  the  $\sigma(T)$ curve may vary at low T between ribbons and ingots of the same RRR. This we attribute to microscopic fluctuations and  difference in microstructure. The  origin of    the   intrinsic insulating behavior of the exemplary i-AlPdRe samples  remains a valid, open and challenging question, which will  require a microscopic knowledge of the samples electronic properties. These results re-open the discussion on the role of quasiperiodic versus periodic order and disorder for electronic propagation.

\appendix

\section{\label{AppendixA} Annealing of i-AlPdRe ribbons}

\begin{figure}[h]
\includegraphics[width=7cm]{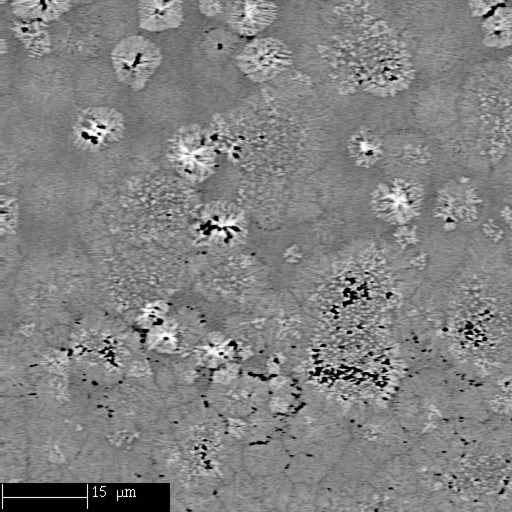}
\includegraphics[width=7cm]{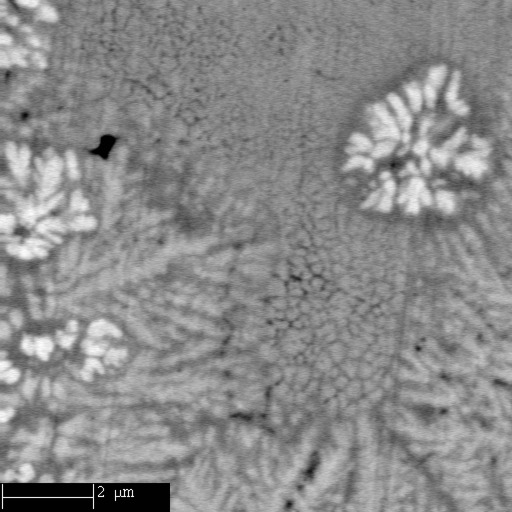}
\caption{\label{FigureA1} SEM picture (backscattered electron mode) of a polished melt-spun ribbon before annealing. Bar scale: $15\mu m$ (top); $2\mu m$ (bottom).}
\end{figure}

After melt-spinning, the ribbons are not homogeneous and contain a large amount of secondary phases (see the SEM picture of Figure \ref{FigureA1}). They are far from a MIT, with room-temperature conductivity in the range $1000 - 3000 \Omega^{-1}cm^{-1}$ and RRR around 1. In this appendix, we give details on how the heat treatment modifies the  chemical homogeneitity of ribbons  and their resistivity.

\begin{figure}[h]
\includegraphics[width=8cm]{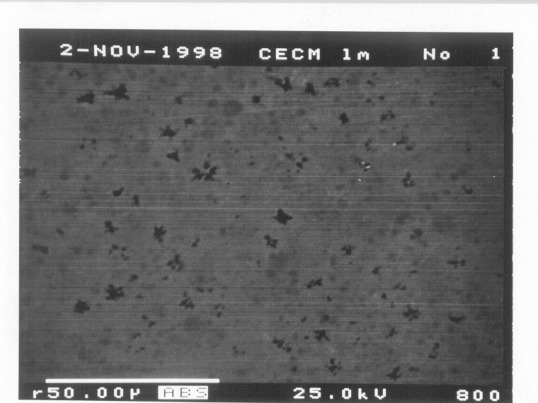}
\includegraphics[width=8cm]{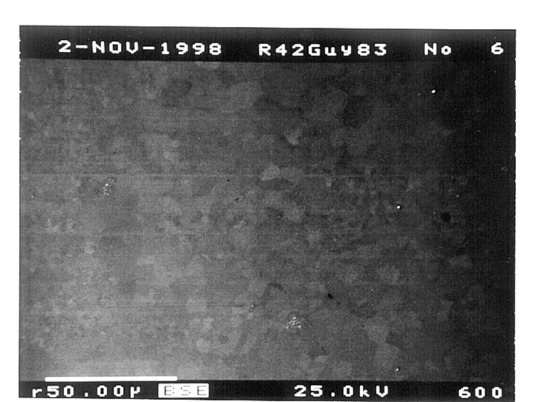}
\caption{\label{FigureA2} Top: SEM image (absorption mode) of a ``low T'' polished melt-spun ribbon annealed 6h at $900^oC$ (the contrast is inverted compared to an image in the backscattering mode). Black phase composition: Al: 71 at.\%, Pd: 12 at.\% and Re: 17 at.\%. Grey phase composition:  Al: 68-70 at.\%, Pd: 20-23.5 at.\% and Re: 8-10.5 at.\%. Bottom: SEM picture (backscattering mode) of a ``high T'' polished melt-spun ribbon annealed 0.1h at $960^oC$. Grains of micrometre size are clearly visible with small contrast from grains to grains which correspond to chemical composition fluctuations of about 1 at.\%}
\end{figure}

We have studied the influence of  annealing parameters on the RRR values of ribbons made by the group of Y. Calvayrac et al (CECM, Vitry, France). The results are listed in Table \ref{TableRecuitCECM}.
For ribbons melt-spun from ingot A, the highest RRR values are obtained after a heat treatment at $900^oC$ during few hours (ribbons ``CECMA1'' and ``CECMA2''). However, the SEM picture on top of Figure \ref{FigureA2} reveals that important inhomogeneities are still present in these ribbons. These ribbons are  called  ``low T'' ribbons in the main text. After  additional heat treatment at a lower temperature of $600^oC$ (``CECMA4'') the RRR slightly decreases, contrary to what was observed in ingots (see subsection \ref{Annealing}). This change in  RRR  with a low T annealing is not reversible: the initial high RRR can not be recovered by a heat treatment at $900^oC$ (``CECMA5''). The ``CECMA3'' ribbons which have been annealed two  hours at $1010^oC$ display the best X-ray diffraction pattern \cite{CalvayracComPrive} and are more homogeneous (see subsection \ref{CompositionOrder}) but do not have the highest RRR. One possible explanation could be that  too long an annealing close to the melting point of the alloy (estimated to be around $1025^oC$), leads to  Al partial evaporation that changes the average composition compared to low T annealed ribbons. Note that the RRR values don't exceed  30 for the ``CECMB'' ribbons melt-spun from ingot B, probably due to a composition difference with ingot A.

\begin{table*}[]
\caption{\label{TableRecuitCECM}
Resistance ratios of melt-spun ribbons as a function of the annealing parameters. When the cooling and heating rates are not specified, they are equal to $100^oC/h$. The ribbons have been obtained from the melt of two different ingots (ingots A and B). Nominal composition: $Al_{70.5}Pd_{21}Re_{8.5}$.}
\begin{tabular}{|c|c|c|c|c|c|}
  \hline
  Sample name  & Ingot & Plateau temperatures and durations & RRR \\ \hline
  CECMA1 & A & $900^oC$ 6h ($50^oC/h$) & 129, 129, 105, 100, 95, 92, 53  \\
  CECMB & B & $900^oC$ 6h  ($50^oC/h$) & 24, 15, 13 \\
  CECMA2 & A & $900^oC$ 6h & 130, 90 \\
  CECMA3 & A & $1010^oC$ 2h & 11, 32, 26, 10, 10 \\
  CECMA4 & A & $900^oC$ 6h + $600^oC$ 2h & 84, 53  \\
  CECMA5 & A & $900^oC$ 6h + $600^oC$ 2h + $900^oC$ 2h + quench & 22, 6, 5  \\ \hline
\end{tabular}
\end{table*}

These results explain the choice of a different heat treatment: a slow heating rate ($100^oC/h$) up to  a short ($\simeq 0.1h$) plateau  at  a temperature just below the melting point ($\simeq 1000^oC$)  to avoid any significant  composition change  and a slow cooling rate ($50-100^oC/h$)  to room temperature. A slow cooling rate should favour a good relaxation of the structure. With such heat treatments, we were able to get (almost) homogeneous ribbons of high structural quality (see  picture at the bottom of Figure \ref{FigureA2}). These ribbons are called  ``high T'' ribbons in the main text . The remaining  composition fluctuations are then of the order of 1 at.\% for the different elements (see subsection \ref{CompositionOrder}).

\begin{figure}[h]
\includegraphics[width=8cm]{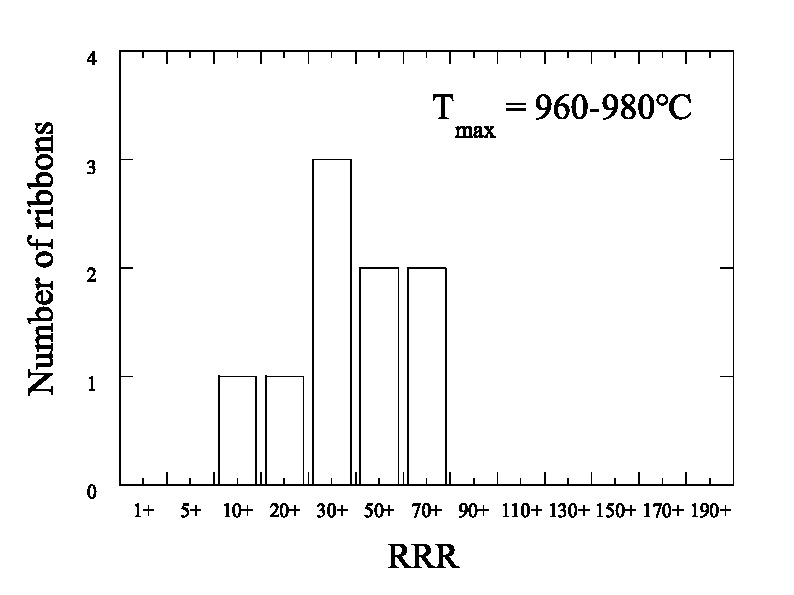}
\includegraphics[width=8cm]{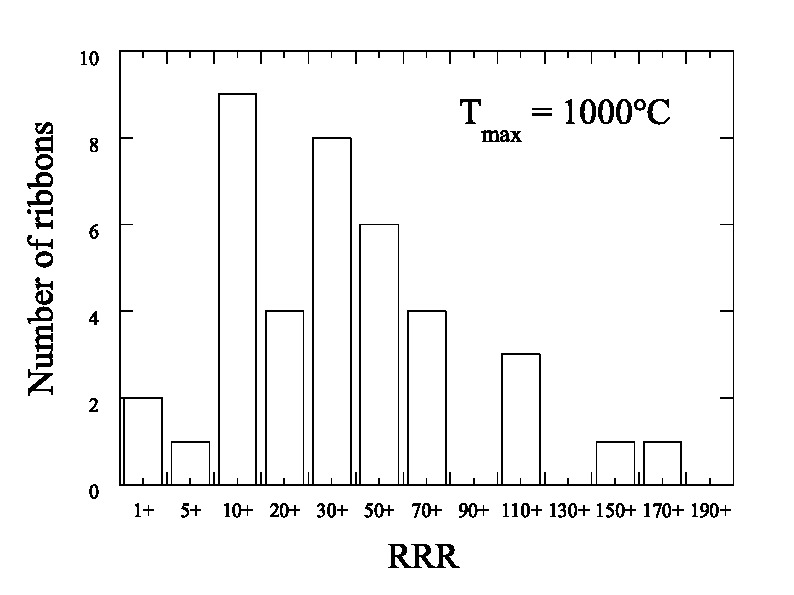}
\includegraphics[width=8cm]{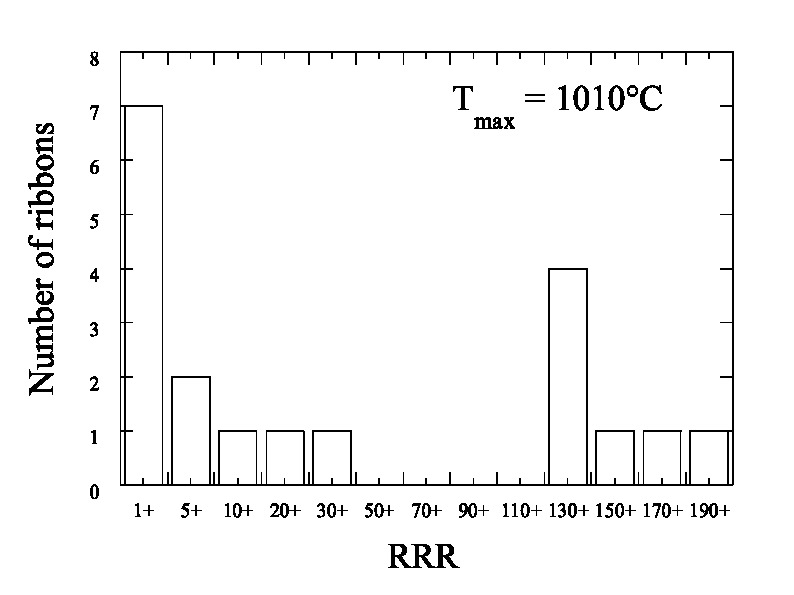}
\caption{\label{FigureA3} Influence of the maximum annealing temperature $T_{max}$ on the RRR values of a set of ``high T'' i-AlPdRe ribbons of nominal composition $Al_{70.5}Pd_{21}Re_{8.5}$. All the ribbons have been melt-spun from the same ingot. From top to bottom: $T_{max}= 960-980^oC$, $1000^oC$ and $1010^oC$.}
\end{figure}

Using the ``high T'' heat treatment described above, we have studied the influence of the maximum annealing temperature $T_{max}$ on the RRR values, all the other annealing parameters  being the same (time spent at $T_{max}$ of 0.1h, heating and cooling rates of $100^oC/h$). For this, many ribbons ($\simeq 1cm$ long) melt-spun from the same ingot were annealed at the same time in a sealed evacuated quartz tube pumped down to high vacuum (base pressure a few $10^{-6} mb$). Histograms of the RRR values are presented in Figure \ref{FigureA3}. First,  samples from the same batch can have broadely different  RRR, between 2 and 200. Second, the $T_{max}$ value strongly influences the
RRR distribution. Annealing at $1010^oC$, very close to the melting point of the alloy, gives RRR below 30 or above 130
 but with no intermediate values. By comparison,  annealing at $960-980^oC$ gives RRR values spread in the range 10 to 90.
 This effect of $T_{max}$ on the RRR distribution may result from composition heterogeneity in the melt before melt-spining that  produces ribbon of various average compositions. The effect of annealing will depend on how close  $T_{max}$ is from  the melting point of the ribbon (that depends on its chemical composition). Interestingly, we have  observed that when no high RRR ribbons can be produced from an  ingot, the ribbons are usually completely molten at $1010^oC$, probably due to a significantly different  composition from high RRR ribbons.
In this regard, we have observed that batches of ribbons of low RRR usually melt at $1010^oC$, probably due to a  composition significantly different  from that of high RRR ribbons.

\section{\label{AppendixB} Is the RRR a good parameter for the ribbons?}

In our i-AlPdRe ribbons, the RRR varies from less than 2 to more than 200, with a  MIT at RRR around 20 \cite{DelahayePRB01}. But how good a parameter is  RRR  to describe the electronic properties of all our ribbons? In other words, is there a one-to-one relation between the RRR values and the observed electronic properties, like the temperature dependence of the conductivity?

\begin{figure}[h]
\includegraphics[width=8cm]{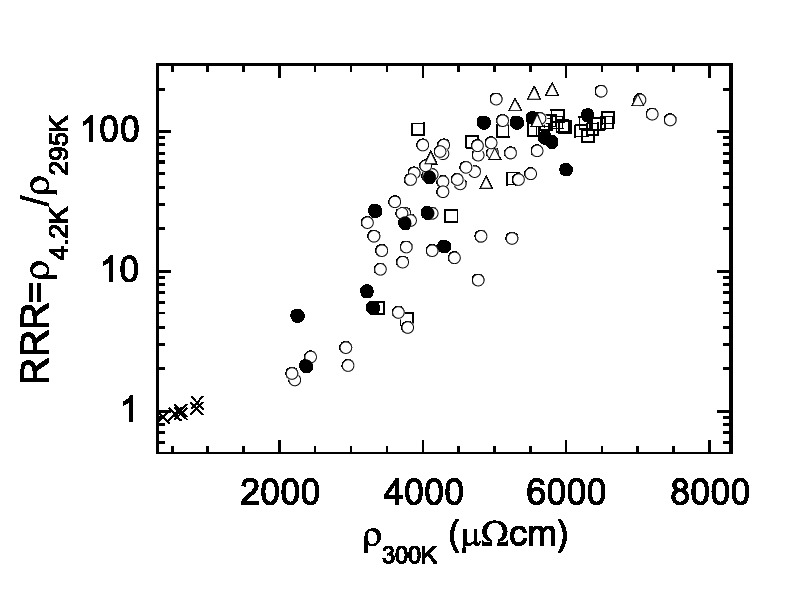}
\caption{\label{FigureB1} RRR values in a log scale as a function of the room temperature resistivity for a large number of ribbons, annealed at different temperatures and melt-spun from different ingots. The x-symbols correspond to non-annealed ribbons.}
\end{figure}

The resistance ratio RRR is, by definition, the resistance (resistivity, resp.) at 4K divided by the resistance (resistivity, resp.) at 300K. This parameter has been extensively used  to categorized the i-AlPdRe samples regarding the MIT. Usually, in disordered systems undergoing a MIT, the driving parameter is known (for example the dopant concentration in doped semi-conductors) and can be used to categorize  the samples. But in i-AlPdRe, the parameter(s) that drive(s) the MIT is (are) not clearly identified and cannot be quantified (see section \ref{PolycrystallineSamples}). To categorize the samples, one may use the room temperature resistivity or conductivity, or  better the low temperature conductivity, which is more sensitive to the approach of the transition.  For our annealed ribbons $\rho_{300K}$ varies between $2000$ and $8000 \Omega cm$ (see Figure \ref{FigureB1}). However in most of  polycrystalline samples, the resistivity can not be determined accurately because of  the porosity of the ingots, the irregular shape of the ribbons, the uncertainties in the geometry of the electrical contacts, etc. In our ribbons, we evaluate the uncertainty in the resistivity or conductivity values to be about 20\%, which is too large to accurately sort out the samples.

The RRR is  a better parameter. It can be determined with a much higher precision (it does not depends on geometrical factors) and it increases rapidly close to the MIT. The RRR was indeed sometimes used also in doped semiconductors since it is much more sensitive than the room temperature resistivity to changes in the dopant concentration  around the MIT (see for example Ref. \cite{CastnerPRB88}).
 The RRR of our ribbons varies by two orders of magnitude (from 2 to 200) and its uncertainty is reasonable  (about 1\%). Within the experimental uncertainties on the resistivity, the RRR  increases almost exponentially with the room temperature resistivity  (see Figure \ref{FigureB1}).

\begin{figure}[h]
\includegraphics[width=8cm]{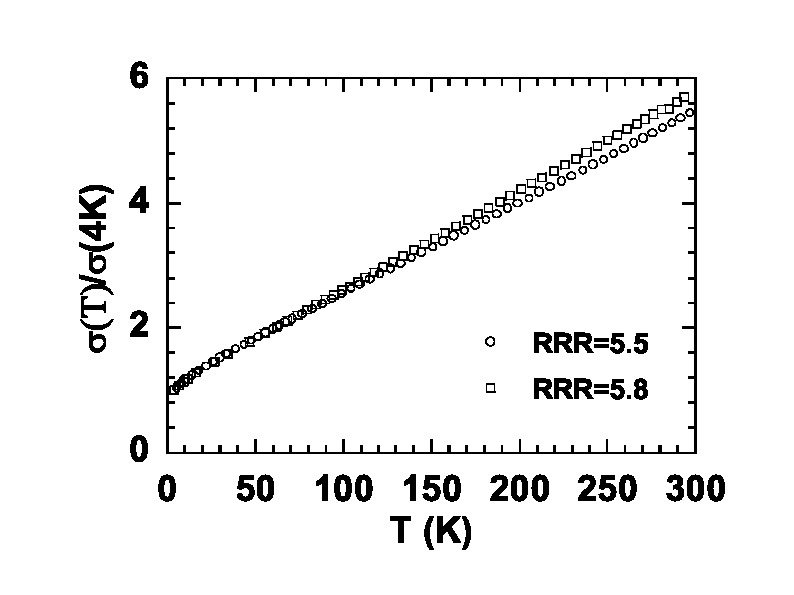}
\caption{\label{FigureB2} Conductivity normalized at 4K as a function of the temperature between 4K and 300K for ribbons of $RRR \simeq 5$ annealed 0.1h at $\simeq 1010^oC$ (``high T'' ribbons).}
\end{figure}

\begin{figure}[h]
\includegraphics[width=8cm]{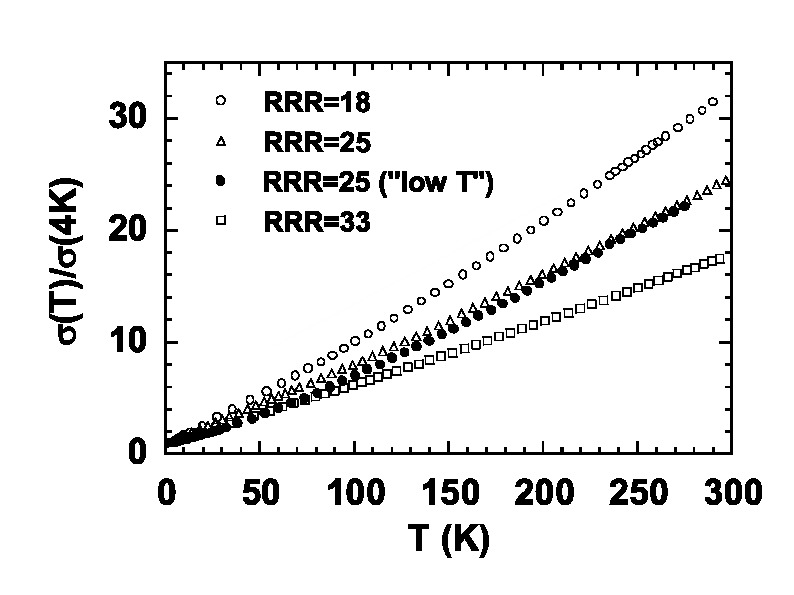}
\includegraphics[width=8cm]{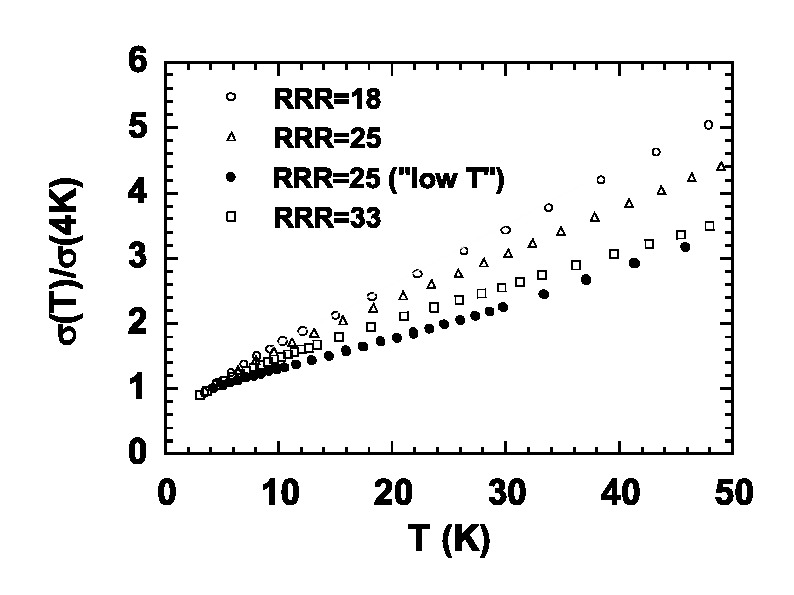}
\caption{\label{FigureB3} Conductivity normalized at 4K as a function of  temperature between 4K and 300K for three ribbons of $ 18 < RRR < 33 $. (Top) T-scale:  0 - 300K. (Bottom) T-scale:  0 - 50K. The ``low T'' ribbon  was annealed at  $900^oC$ whereas the  other three (``high T'') were annealed close to $1000^oC$.}
\end{figure}

\begin{figure}[h]
\includegraphics[width=8cm]{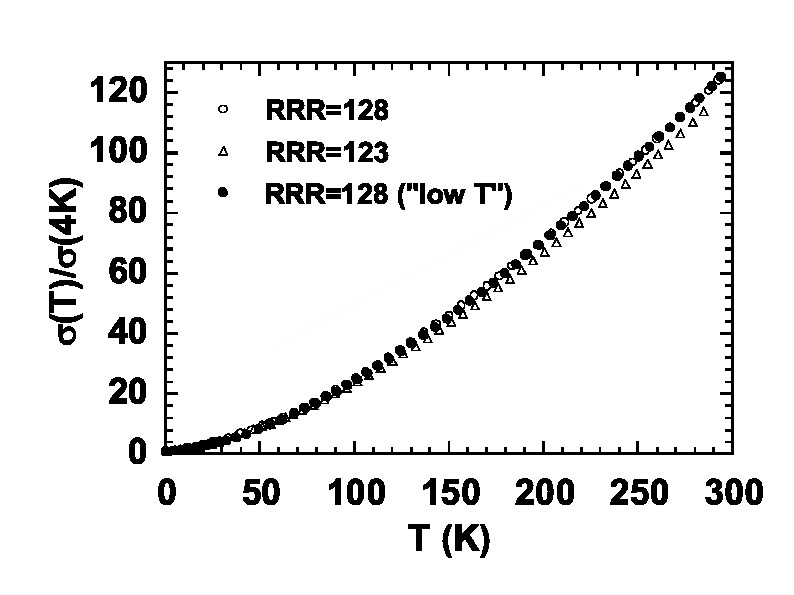}
\includegraphics[width=8cm]{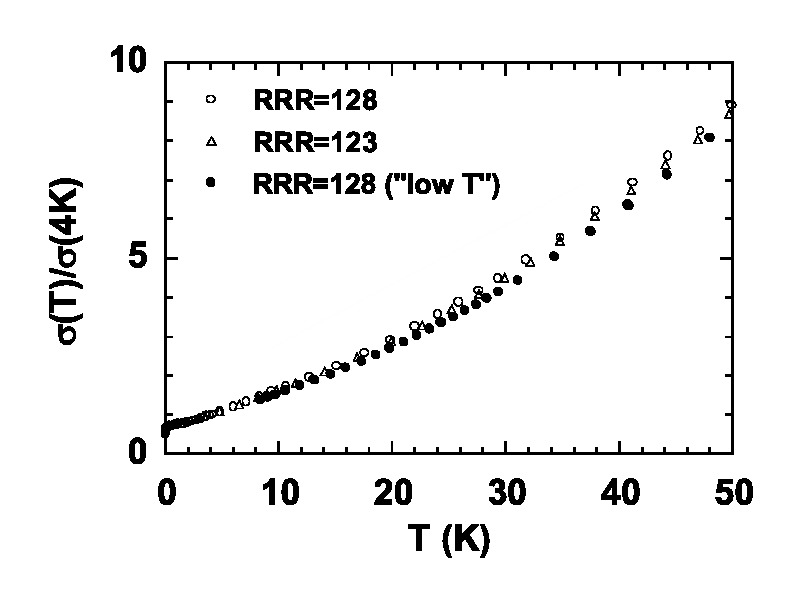}
\caption{\label{FigureB4} Conductivity normalized at 4K as a function of  temperature between 4K and 300K for ribbons of $RRR \simeq 130$ (the ``high T'' ribbon with RRR = 128 was measured down to 20mK). (Top) T-scale:  0 - 300K. (Bottom) T-scale:  0 - 50K The ``low T'' ribbon has been annealed at only $900^oC$ whereas the two other (``high T'' ribbons) have been annealed close to $1000^oC$.}
\end{figure}

  The next  question is  whether there is a strictly one to one correspondence between the RRR values and the $\sigma(T)$ curve.
Most $\sigma(T)$ curves plotted in Figures \ref{FigureB2}, \ref{FigureB3} and \ref{FigureB4} display a steady evolution with the RRR, as reported previously \cite{AhlgrenPRB97,DelahayePRB01,DelahayeJPCM03}, but this is not always the case. For example, in Figure \ref{FigureB3}, a ribbon of RRR = 25 has an ``anomalous'' temperature dependence compared to two other ribbons of similar RRR. The difference is more pronounced at low temperature. The same is observed for another ribbon of RRR = 129 which $\sigma(T)$ is not  intermediate  between two ribbons of RRR = 128 and 130 (Figure \ref{FigureB4}). Its $\sigma(T)$ curve is identical to the ribbon of  RRR = 128  above $\simeq 200K$ and  starts to deviate at lower temperatures. Interestingly, both  ``anomalous'' ribbons have  been annealed to only $900^o$ for 6h (``low T'' ribbons), whereas all the other ribbons have been annealed to about $1000^oC$ for 0.1h (``high T'' ribbons).

\begin{figure}[h]
\includegraphics[width=8cm]{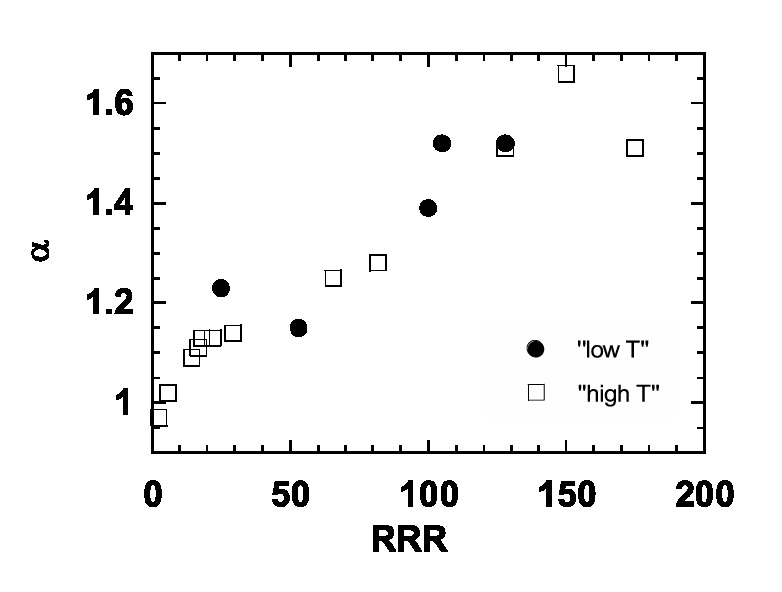}
\caption{\label{FigureB5} Power law $\alpha$ coefficient as a function of  RRR for i-AlPdRe ribbons ($\alpha$  is determined from the fit   $\sigma(T) = \sigma_0+\sigma_1T^{\alpha}$). Open squares: ``high T'' ribbons. Filled circles: ``low T'' ribbons. The uncertainties are of about 5\% for $\alpha$ and few \% for the RRR values.}
\end{figure}

The non universality of the $\sigma(T)$ behavior at low T in our i-AlPdRe ribbons is also exemplified in Figure \ref{FigureB5}. Above about $30K$ up to at least $300K$ , the temperature dependence of the ribbons conductivity is well described by a power law \cite{DelahayeJPCM03}: $\sigma(T) = \sigma_0+\sigma_1T^{\alpha}$.  The $\alpha$ values plotted in Figure \ref{FigureB5} increase from 1 to about 1.5 as a function of the RRR spaning a range 2 to 200. The scattering in the data reflects the absence of an universal behavior for all the ribbons. It is clearly more pronounced for the ``low T'' ribbons, but an extreme value of 1.66 is also observed in a ``high T'' ribbon, indicating that the anomalous behavior is not limited to the ``low T'' ribbons.

\begin{figure}[h]
\includegraphics[width=8cm]{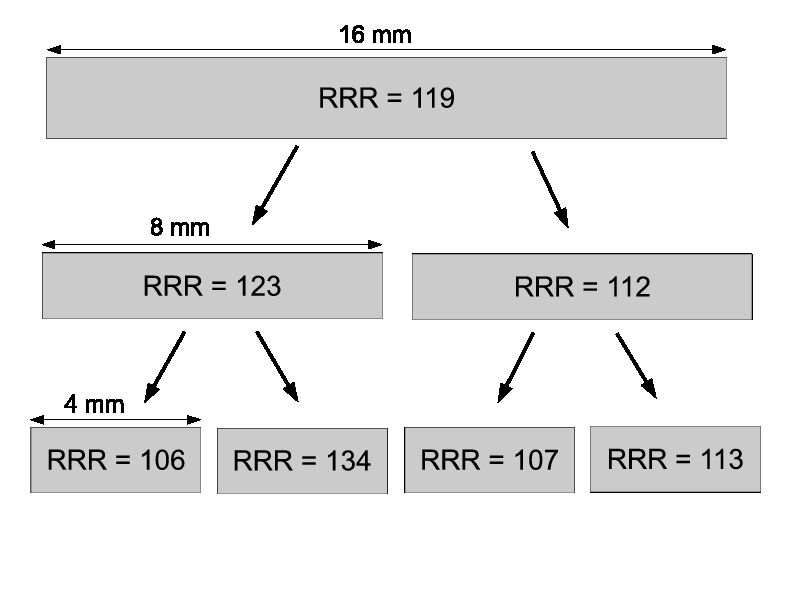}
\caption{\label{FigureB6} RRR values measured on a ribbon 16mm long cut in two and four pieces of identical sizes (``high T'' ribbon).}
\end{figure}

We have shown (Appendix \ref{AppendixA} and subsection \ref{CompositionOrder}) that the ``low T'' ribbons annealed at $900^oC$ have  larger spatial  composition fluctuations  than the ``high T'' ribbons annealed closer to the alloy melting point. The anomalous $\sigma(T)$ behaviors appear therefore  related to larger chemical homogeneity in the ribbons.  We demonstrate in Figure \ref{FigureB6} the existence of electrical inhomogeneities in highly resistive ribbons. We have cut  a ribbon of RRR = 119 into four pieces of the same size. The ribbon was 16mm long and had been annealed at $\simeq 1000^oC$. The RRR of each pieces is between 106 and 134. We canot exclude larger fluctuations on a smaller length scale. This simple experiment shows that the RRR value measured on a ribbon is the result of some complex electrical combinations of smaller size domains that have different RRR values. Differences in the amplitudes or the typical length scales of these electrical inhomogeneities can obviously give differences in $\sigma(T)$ curves.

\begin{figure}[h]
\includegraphics[width=8cm]{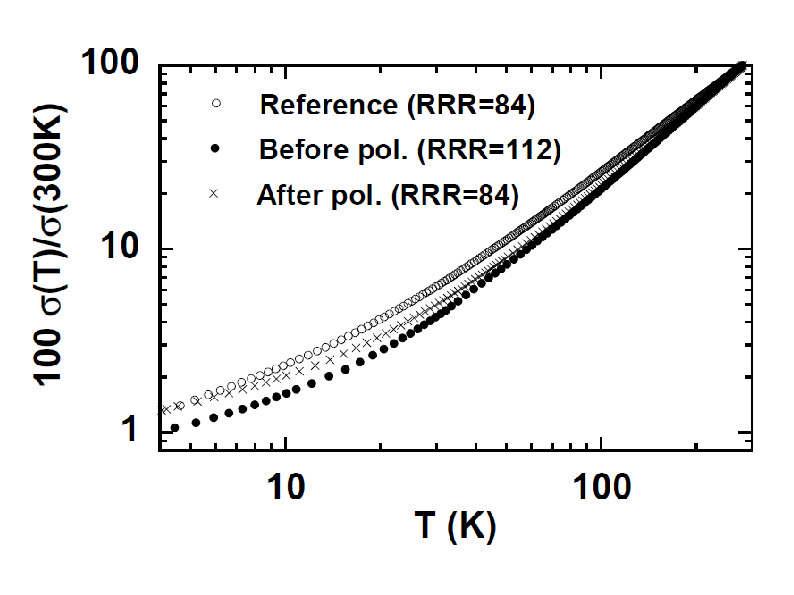}
\caption{\label{FigureB7} Normalized $\sigma(T)$ curves measured on a ribbon before (RRR = 112) and after (RRR = 84) polishing. The $\sigma(T)$ curve of a (Reference) non polished ribbon with RRR = 84  is also plotted for comparison.}
\end{figure}

We have also studied the electrical homogeneity in the thickness of the ribbons by measuring the $\sigma(T)$ curves on two ribbons of $RRR \simeq 120$  before and after one third of the film thickness was removed by polishing. The RRR was always smaller (RRR=60-90) after polishing than before (RRR=110-120). The typical evolution of the $\sigma(T)$ curves is shown in Figure \ref{FigureB7}. The high temperature behavior (above 50K) is almost unchanged by polishing while the conductivity variations are less pronounced at low temperature. The $\sigma(T)$ curves of the polished ribbons are also anomalous compared to non polished ones of similar RRR. Wether this result is related to the small chemical gradient, often observed in the thickness of the ribbons, or if it is due a disordered and more conducting layer induced at the surface of the ribbons by the mechanical polishing process remains an open question.

 \clearpage

\begin{acknowledgments}

We would like to acknowledge Thierry Grenet for fruitful comments and discussions. We thank Guy Fourcaudot, Jean-Claude Grieco and Yvonne Calvayrac for the ribbons elaboration, Pierre Amiot and Fran\c coise Robaut for the SEM and microprobe analysis, and Marie-Claude Cheynet for the TEM picture of a granular Al film.  Osten Rapp and Markus Rodmar (KTH, Stockholm) are thanked for the measurements of the ribbons presented in Fig. 6.

\end{acknowledgments}


\providecommand{\noopsort}[1]{}\providecommand{\singleletter}[1]{#1}%

%


\begin{thebibliography}{60}%
\makeatletter
\providecommand \@ifxundefined [1]{%
 \@ifx{#1\undefined}
}%
\providecommand \@ifnum [1]{%
 \ifnum #1\expandafter \@firstoftwo
 \else \expandafter \@secondoftwo
 \fi
}%
\providecommand \@ifx [1]{%
 \ifx #1\expandafter \@firstoftwo
 \else \expandafter \@secondoftwo
 \fi
}%
\providecommand \natexlab [1]{#1}%
\providecommand \enquote  [1]{``#1''}%
\providecommand \bibnamefont  [1]{#1}%
\providecommand \bibfnamefont [1]{#1}%
\providecommand \citenamefont [1]{#1}%
\providecommand \href@noop [0]{\@secondoftwo}%
\providecommand \href [0]{\begingroup \@sanitize@url \@href}%
\providecommand \@href[1]{\@@startlink{#1}\@@href}%
\providecommand \@@href[1]{\endgroup#1\@@endlink}%
\providecommand \@sanitize@url [0]{\catcode `\\12\catcode `\$12\catcode
  `\&12\catcode `\#12\catcode `\^12\catcode `\_12\catcode `\%12\relax}%
\providecommand \@@startlink[1]{}%
\providecommand \@@endlink[0]{}%
\providecommand \url  [0]{\begingroup\@sanitize@url \@url }%
\providecommand \@url [1]{\endgroup\@href {#1}{\urlprefix }}%
\providecommand \urlprefix  [0]{URL }%
\providecommand \Eprint [0]{\href }%
\providecommand \doibase [0]{http://dx.doi.org/}%
\providecommand \selectlanguage [0]{\@gobble}%
\providecommand \bibinfo  [0]{\@secondoftwo}%
\providecommand \bibfield  [0]{\@secondoftwo}%
\providecommand \translation [1]{[#1]}%
\providecommand \BibitemOpen [0]{}%
\providecommand \bibitemStop [0]{}%
\providecommand \bibitemNoStop [0]{.\EOS\space}%
\providecommand \EOS [0]{\spacefactor3000\relax}%
\providecommand \BibitemShut  [1]{\csname bibitem#1\endcsname}%
\let\auto@bib@innerbib\@empty
\bibitem [{Note1()}]{Note1}%
  \BibitemOpen
  \bibinfo {note} {An insulator is by definition a material with a zero
  conductivity (infinite resistivity) at zero temperature, while the
  conductivity remains non-zero (finite resistivity) in a metallic
  system.}\BibitemShut {Stop}%
\bibitem [{\citenamefont {Rapp}(1999)}]{RappBook99}%
  \BibitemOpen
  \bibfield  {author} {\bibinfo {author} {\bibfnamefont {O.}~\bibnamefont
  {Rapp}},\ }\enquote {\bibinfo {title} {Physical properties of
  quasicrystals},}\ \ (\bibinfo  {publisher} {Springer},\ \bibinfo {year}
  {1999})\ Chap.~\bibinfo {chapter} {5}, pp.\ \bibinfo {pages}
  {127--167}\BibitemShut {NoStop}%
\bibitem [{\citenamefont {Poon}\ \emph {et~al.}(1998)\citenamefont {Poon},
  \citenamefont {Zavaliche},\ and\ \citenamefont {Beeli}}]{PoonMRS99}%
  \BibitemOpen
  \bibfield  {author} {\bibinfo {author} {\bibfnamefont {S.~J.}\ \bibnamefont
  {Poon}}, \bibinfo {author} {\bibfnamefont {F.}~\bibnamefont {Zavaliche}}, \
  and\ \bibinfo {author} {\bibfnamefont {C.}~\bibnamefont {Beeli}},\
  }\href@noop {} {\bibfield  {journal} {\bibinfo  {journal} {MRS Proceedings}\
  }\textbf {\bibinfo {volume} {553}},\ \bibinfo {pages} {365} (\bibinfo {year}
  {1998})}\BibitemShut {NoStop}%
\bibitem [{\citenamefont {Delahaye}\ \emph
  {et~al.}(2003{\natexlab{a}})\citenamefont {Delahaye}, \citenamefont
  {Berger},\ and\ \citenamefont {Fourcaudot}}]{DelahayeJPCM03}%
  \BibitemOpen
  \bibfield  {author} {\bibinfo {author} {\bibfnamefont {J.}~\bibnamefont
  {Delahaye}}, \bibinfo {author} {\bibfnamefont {C.}~\bibnamefont {Berger}}, \
  and\ \bibinfo {author} {\bibfnamefont {G.}~\bibnamefont {Fourcaudot}},\
  }\href@noop {} {\bibfield  {journal} {\bibinfo  {journal} {J. Phys.: Condens.
  Matter}\ }\textbf {\bibinfo {volume} {15}},\ \bibinfo {pages} {8753}
  (\bibinfo {year} {2003}{\natexlab{a}})}\BibitemShut {NoStop}%
\bibitem [{\citenamefont {Guo}\ \emph {et~al.}(2000)\citenamefont {Guo},
  \citenamefont {Sato}, \citenamefont {Abe}, \citenamefont {Takakura},\ and\
  \citenamefont {Tsai}}]{GuoPML00}%
  \BibitemOpen
  \bibfield  {author} {\bibinfo {author} {\bibfnamefont {J.~Q.}\ \bibnamefont
  {Guo}}, \bibinfo {author} {\bibfnamefont {T.~J.}\ \bibnamefont {Sato}},
  \bibinfo {author} {\bibfnamefont {E.}~\bibnamefont {Abe}}, \bibinfo {author}
  {\bibfnamefont {H.}~\bibnamefont {Takakura}}, \ and\ \bibinfo {author}
  {\bibfnamefont {A.~P.}\ \bibnamefont {Tsai}},\ }\href@noop {} {\bibfield
  {journal} {\bibinfo  {journal} {Phil. Mag. Lett.}\ }\textbf {\bibinfo
  {volume} {80}},\ \bibinfo {pages} {495} (\bibinfo {year} {2000})}\BibitemShut
  {NoStop}%
\bibitem [{\citenamefont {Fisher}\ \emph {et~al.}(2002)\citenamefont {Fisher},
  \citenamefont {Xie}, \citenamefont {Tudosa}, \citenamefont {Gao},
  \citenamefont {Song}, \citenamefont {Canfield}, \citenamefont {Kracher},
  \citenamefont {Dennis}, \citenamefont {Abanoz},\ and\ \citenamefont
  {Kramer}}]{FisherPMB02}%
  \BibitemOpen
  \bibfield  {author} {\bibinfo {author} {\bibfnamefont {I.~R.}\ \bibnamefont
  {Fisher}}, \bibinfo {author} {\bibfnamefont {X.~P.}\ \bibnamefont {Xie}},
  \bibinfo {author} {\bibfnamefont {I.}~\bibnamefont {Tudosa}}, \bibinfo
  {author} {\bibfnamefont {C.~W.}\ \bibnamefont {Gao}}, \bibinfo {author}
  {\bibfnamefont {C.}~\bibnamefont {Song}}, \bibinfo {author} {\bibfnamefont
  {P.~C.}\ \bibnamefont {Canfield}}, \bibinfo {author} {\bibfnamefont
  {A.}~\bibnamefont {Kracher}}, \bibinfo {author} {\bibfnamefont
  {K.}~\bibnamefont {Dennis}}, \bibinfo {author} {\bibfnamefont
  {D.}~\bibnamefont {Abanoz}}, \ and\ \bibinfo {author} {\bibfnamefont {M.~J.}\
  \bibnamefont {Kramer}},\ }\href@noop {} {\bibfield  {journal} {\bibinfo
  {journal} {Phil. Mag. Lett.}\ }\textbf {\bibinfo {volume} {82}},\ \bibinfo
  {pages} {1089} (\bibinfo {year} {2002})}\BibitemShut {NoStop}%
\bibitem [{\citenamefont {Dolin\v{s}ek}\ \emph {et~al.}(2006)\citenamefont
  {Dolin\v{s}ek}, \citenamefont {McGuiness}, \citenamefont {Jlanj\v{s}ek},
  \citenamefont {Smiljani\'c}, \citenamefont {Smontara}, \citenamefont
  {Zijlstra}, \citenamefont {Bose}, \citenamefont {Fisher}, \citenamefont
  {Kramer},\ and\ \citenamefont {Canfield}}]{DolinsekPRB06}%
  \BibitemOpen
  \bibfield  {author} {\bibinfo {author} {\bibfnamefont {J.}~\bibnamefont
  {Dolin\v{s}ek}}, \bibinfo {author} {\bibfnamefont {P.~J.}\ \bibnamefont
  {McGuiness}}, \bibinfo {author} {\bibfnamefont {M.}~\bibnamefont
  {Jlanj\v{s}ek}}, \bibinfo {author} {\bibfnamefont {I.}~\bibnamefont
  {Smiljani\'c}}, \bibinfo {author} {\bibfnamefont {A.}~\bibnamefont
  {Smontara}}, \bibinfo {author} {\bibfnamefont {E.~S.}\ \bibnamefont
  {Zijlstra}}, \bibinfo {author} {\bibfnamefont {S.~K.}\ \bibnamefont {Bose}},
  \bibinfo {author} {\bibfnamefont {I.~R.}\ \bibnamefont {Fisher}}, \bibinfo
  {author} {\bibfnamefont {M.~J.}\ \bibnamefont {Kramer}}, \ and\ \bibinfo
  {author} {\bibfnamefont {P.~C.}\ \bibnamefont {Canfield}},\ }\href@noop {}
  {\bibfield  {journal} {\bibinfo  {journal} {Phys. Rev. B}\ }\textbf {\bibinfo
  {volume} {74}},\ \bibinfo {pages} {134201} (\bibinfo {year}
  {2006})}\BibitemShut {NoStop}%
\bibitem [{\citenamefont {Vekilov}\ and\ \citenamefont
  {Chernikov}(2009)}]{VekilovEPL09}%
  \BibitemOpen
  \bibfield  {author} {\bibinfo {author} {\bibfnamefont {Y.~K.}\ \bibnamefont
  {Vekilov}}\ and\ \bibinfo {author} {\bibfnamefont {M.~A.}\ \bibnamefont
  {Chernikov}},\ }\href@noop {} {\bibfield  {journal} {\bibinfo  {journal}
  {Eur. Phys. Lett.}\ }\textbf {\bibinfo {volume} {87}},\ \bibinfo {pages}
  {17010} (\bibinfo {year} {2009})}\BibitemShut {NoStop}%
\bibitem [{\citenamefont {Poon}\ and\ \citenamefont {Rapp}(2007)}]{PoonPRB07}%
  \BibitemOpen
  \bibfield  {author} {\bibinfo {author} {\bibfnamefont {S.~J.}\ \bibnamefont
  {Poon}}\ and\ \bibinfo {author} {\bibfnamefont {O.}~\bibnamefont {Rapp}},\
  }\href@noop {} {\bibfield  {journal} {\bibinfo  {journal} {Phys. Rev. B}\
  }\textbf {\bibinfo {volume} {76}},\ \bibinfo {pages} {216201} (\bibinfo
  {year} {2007})}\BibitemShut {NoStop}%
\bibitem [{\citenamefont {Rapp}\ and\ \citenamefont {Poon}(2011)}]{RappPRB11}%
  \BibitemOpen
  \bibfield  {author} {\bibinfo {author} {\bibfnamefont {O.}~\bibnamefont
  {Rapp}}\ and\ \bibinfo {author} {\bibfnamefont {S.~J.}\ \bibnamefont
  {Poon}},\ }\href@noop {} {\bibfield  {journal} {\bibinfo  {journal} {Phys.
  Rev. B}\ }\textbf {\bibinfo {volume} {84}},\ \bibinfo {pages} {174206}
  (\bibinfo {year} {2011})}\BibitemShut {NoStop}%
\bibitem [{\citenamefont {Dolin\ifmmode~\check{s}\else \v{s}\fi{}ek}\ \emph
  {et~al.}(2007)\citenamefont {Dolin\ifmmode~\check{s}\else \v{s}\fi{}ek},
  \citenamefont {McGuiness}, \citenamefont {Klanj\ifmmode~\check{s}\else
  \v{s}\fi{}ek}, \citenamefont {Smiljani\ifmmode~\acute{c}\else \'{c}\fi{}},
  \citenamefont {Smontara}, \citenamefont {Zijlstra}, \citenamefont {Bose},
  \citenamefont {Fisher}, \citenamefont {Kramer},\ and\ \citenamefont
  {Canfield}}]{DolinsekPRB07}%
  \BibitemOpen
  \bibfield  {author} {\bibinfo {author} {\bibfnamefont {J.}~\bibnamefont
  {Dolin\ifmmode~\check{s}\else \v{s}\fi{}ek}}, \bibinfo {author}
  {\bibfnamefont {P.~J.}\ \bibnamefont {McGuiness}}, \bibinfo {author}
  {\bibfnamefont {M.}~\bibnamefont {Klanj\ifmmode~\check{s}\else
  \v{s}\fi{}ek}}, \bibinfo {author} {\bibfnamefont {I.}~\bibnamefont
  {Smiljani\ifmmode~\acute{c}\else \'{c}\fi{}}}, \bibinfo {author}
  {\bibfnamefont {A.}~\bibnamefont {Smontara}}, \bibinfo {author}
  {\bibfnamefont {E.~S.}\ \bibnamefont {Zijlstra}}, \bibinfo {author}
  {\bibfnamefont {S.~K.}\ \bibnamefont {Bose}}, \bibinfo {author}
  {\bibfnamefont {I.~R.}\ \bibnamefont {Fisher}}, \bibinfo {author}
  {\bibfnamefont {M.~J.}\ \bibnamefont {Kramer}}, \ and\ \bibinfo {author}
  {\bibfnamefont {P.~C.}\ \bibnamefont {Canfield}},\ }\href {\doibase
  10.1103/PhysRevB.76.216202} {\bibfield  {journal} {\bibinfo  {journal} {Phys.
  Rev. B}\ }\textbf {\bibinfo {volume} {76}},\ \bibinfo {pages} {216202}
  (\bibinfo {year} {2007})}\BibitemShut {NoStop}%
\bibitem [{\citenamefont {Delahaye}(2000)}]{DelahayeThesis99}%
  \BibitemOpen
  \bibfield  {author} {\bibinfo {author} {\bibfnamefont {J.}~\bibnamefont
  {Delahaye}},\ }\emph {\bibinfo {title} {Etude Exp\'erimentale de la
  Transition M\'etal-Isolant dans les Quasicristaux}},\ \href@noop {} {\bibinfo
  {type} {{Ph.D.} thesis}},\ \bibinfo  {school} {Universit\'e Joseph Fourier
  Grenoble} (\bibinfo {year} {2000})\BibitemShut {NoStop}%
\bibitem [{\citenamefont {Honda}\ \emph {et~al.}(1994)\citenamefont {Honda},
  \citenamefont {Edagawa}, \citenamefont {Yoshioka}, \citenamefont
  {Hashimoto},\ and\ \citenamefont {Takeuchi}}]{HondaJJAP94}%
  \BibitemOpen
  \bibfield  {author} {\bibinfo {author} {\bibfnamefont {Y.}~\bibnamefont
  {Honda}}, \bibinfo {author} {\bibfnamefont {K.}~\bibnamefont {Edagawa}},
  \bibinfo {author} {\bibfnamefont {A.}~\bibnamefont {Yoshioka}}, \bibinfo
  {author} {\bibfnamefont {T.}~\bibnamefont {Hashimoto}}, \ and\ \bibinfo
  {author} {\bibfnamefont {S.}~\bibnamefont {Takeuchi}},\ }\href@noop {}
  {\bibfield  {journal} {\bibinfo  {journal} {Jpn. J. Appl. Phys.}\ }\textbf
  {\bibinfo {volume} {33}},\ \bibinfo {pages} {4929} (\bibinfo {year}
  {1994})}\BibitemShut {NoStop}%
\bibitem [{\citenamefont {Pierce}\ \emph {et~al.}(1994)\citenamefont {Pierce},
  \citenamefont {Guo},\ and\ \citenamefont {Poon}}]{PiercePRL94}%
  \BibitemOpen
  \bibfield  {author} {\bibinfo {author} {\bibfnamefont {F.~S.}\ \bibnamefont
  {Pierce}}, \bibinfo {author} {\bibfnamefont {Q.}~\bibnamefont {Guo}}, \ and\
  \bibinfo {author} {\bibfnamefont {S.~J.}\ \bibnamefont {Poon}},\ }\href@noop
  {} {\bibfield  {journal} {\bibinfo  {journal} {Phys. Rev. Lett.}\ }\textbf
  {\bibinfo {volume} {73}},\ \bibinfo {pages} {2220} (\bibinfo {year}
  {1994})}\BibitemShut {NoStop}%
\bibitem [{\citenamefont {Sawada}\ \emph {et~al.}(1998)\citenamefont {Sawada},
  \citenamefont {Tamura}, \citenamefont {Kimura},\ and\ \citenamefont
  {Ino}}]{SawadaICQ6}%
  \BibitemOpen
  \bibfield  {author} {\bibinfo {author} {\bibfnamefont {H.}~\bibnamefont
  {Sawada}}, \bibinfo {author} {\bibfnamefont {R.}~\bibnamefont {Tamura}},
  \bibinfo {author} {\bibfnamefont {K.}~\bibnamefont {Kimura}}, \ and\ \bibinfo
  {author} {\bibfnamefont {H.}~\bibnamefont {Ino}},\ }in\ \href@noop {} {\emph
  {\bibinfo {booktitle} {Proceedings of the 6th International Conference on
  Quasicrystals (ICQ6)}}},\ \bibinfo {editor} {edited by\ \bibinfo {editor}
  {\bibfnamefont {S.}~\bibnamefont {Takeuchi}}\ and\ \bibinfo {editor}
  {\bibfnamefont {T.}~\bibnamefont {Fujiwara}}}\ (\bibinfo  {publisher} {World
  Scientific},\ \bibinfo {year} {1998})\ p.\ \bibinfo {pages} {329}\BibitemShut
  {NoStop}%
\bibitem [{\citenamefont {Rosenbaum}\ \emph {et~al.}(2004)\citenamefont
  {Rosenbaum}, \citenamefont {Murphy}, \citenamefont {Brandt}, \citenamefont
  {Wang}, \citenamefont {Zhong}, \citenamefont {Wu}, \citenamefont {Lin},\ and\
  \citenamefont {Lin}}]{RosenbaumJPCM04}%
  \BibitemOpen
  \bibfield  {author} {\bibinfo {author} {\bibfnamefont {R.}~\bibnamefont
  {Rosenbaum}}, \bibinfo {author} {\bibfnamefont {T.}~\bibnamefont {Murphy}},
  \bibinfo {author} {\bibfnamefont {B.}~\bibnamefont {Brandt}}, \bibinfo
  {author} {\bibfnamefont {C.~R.}\ \bibnamefont {Wang}}, \bibinfo {author}
  {\bibfnamefont {Y.~L.}\ \bibnamefont {Zhong}}, \bibinfo {author}
  {\bibfnamefont {S.~W.}\ \bibnamefont {Wu}}, \bibinfo {author} {\bibfnamefont
  {S.~T.}\ \bibnamefont {Lin}}, \ and\ \bibinfo {author} {\bibfnamefont
  {J.~J.}\ \bibnamefont {Lin}},\ }\href@noop {} {\bibfield  {journal} {\bibinfo
   {journal} {J. Phys.: Condens. Matter}\ }\textbf {\bibinfo {volume} {16}},\
  \bibinfo {pages} {821} (\bibinfo {year} {2004})}\BibitemShut {NoStop}%
\bibitem [{\citenamefont {Bianchi}\ \emph {et~al.}(1997)\citenamefont
  {Bianchi}, \citenamefont {Bommeli}, \citenamefont {Chernikov}, \citenamefont
  {Gubler}, \citenamefont {Degiorgi},\ and\ \citenamefont
  {Ott}}]{BianchiPRB97}%
  \BibitemOpen
  \bibfield  {author} {\bibinfo {author} {\bibfnamefont {A.~D.}\ \bibnamefont
  {Bianchi}}, \bibinfo {author} {\bibfnamefont {F.}~\bibnamefont {Bommeli}},
  \bibinfo {author} {\bibfnamefont {M.~A.}\ \bibnamefont {Chernikov}}, \bibinfo
  {author} {\bibfnamefont {U.}~\bibnamefont {Gubler}}, \bibinfo {author}
  {\bibfnamefont {L.}~\bibnamefont {Degiorgi}}, \ and\ \bibinfo {author}
  {\bibfnamefont {H.~R.}\ \bibnamefont {Ott}},\ }\href@noop {} {\bibfield
  {journal} {\bibinfo  {journal} {Phys. Rev. B}\ }\textbf {\bibinfo {volume}
  {55}},\ \bibinfo {pages} {5730} (\bibinfo {year} {1997})}\BibitemShut
  {NoStop}%
\bibitem [{\citenamefont {Berger}\ \emph {et~al.}(1993)\citenamefont {Berger},
  \citenamefont {Grenet}, \citenamefont {Lindqvist}, \citenamefont {Lanco},
  \citenamefont {Grieco}, \citenamefont {Fourcaudot},\ and\ \citenamefont
  {Cyrot-Lackmann}}]{BergerSSC93}%
  \BibitemOpen
  \bibfield  {author} {\bibinfo {author} {\bibfnamefont {C.}~\bibnamefont
  {Berger}}, \bibinfo {author} {\bibfnamefont {T.}~\bibnamefont {Grenet}},
  \bibinfo {author} {\bibfnamefont {P.}~\bibnamefont {Lindqvist}}, \bibinfo
  {author} {\bibfnamefont {P.}~\bibnamefont {Lanco}}, \bibinfo {author}
  {\bibfnamefont {J.~C.}\ \bibnamefont {Grieco}}, \bibinfo {author}
  {\bibfnamefont {G.}~\bibnamefont {Fourcaudot}}, \ and\ \bibinfo {author}
  {\bibfnamefont {F.}~\bibnamefont {Cyrot-Lackmann}},\ }\href@noop {}
  {\bibfield  {journal} {\bibinfo  {journal} {Solid State Communications}\
  }\textbf {\bibinfo {volume} {87}},\ \bibinfo {pages} {977} (\bibinfo {year}
  {1993})}\BibitemShut {NoStop}%
\bibitem [{\citenamefont {Delahaye}(1997)}]{DelahayeDEA97}%
  \BibitemOpen
  \bibfield  {author} {\bibinfo {author} {\bibfnamefont {J.}~\bibnamefont
  {Delahaye}},\ }\emph {\bibinfo {title} {Mesures de transport et transition
  m\'etal/isolant dans le syst\`eme ordonn\'e quasicristallin AlPdRe}},\
  \href@noop {} {\bibinfo {type} {Dea report}},\ \bibinfo  {school}
  {Universit\'e Joseph Fourier Grenoble} (\bibinfo {year} {1997})\BibitemShut
  {NoStop}%
\bibitem [{\citenamefont {Haberkern}(1999)}]{HaberkernCondmat99}%
  \BibitemOpen
  \bibfield  {author} {\bibinfo {author} {\bibfnamefont {R.}~\bibnamefont
  {Haberkern}},\ }\href@noop {} {\enquote {\bibinfo {title} {Electronic
  transport properties of quasicrystalline thin films},}\ }\bibinfo
  {howpublished} {e-print arXiv:cond-mat/9911426} (\bibinfo {year} {1999}),\
  \bibinfo {note} {tutorial review presented at a summer school on
  quasicrystals in Chemnitz 1997}\BibitemShut {NoStop}%
\bibitem [{\citenamefont {Bergman}(1999)}]{BergmanStage99}%
  \BibitemOpen
  \bibfield  {author} {\bibinfo {author} {\bibfnamefont {A.}~\bibnamefont
  {Bergman}},\ }\emph {\bibinfo {title} {Elaboration et Charact\'erisation
  Electrique de Couches Minces du Quasicristal i-AlPdRe}},\ \href@noop {}
  {\bibinfo {type} {Final study report}},\ \bibinfo  {school} {LEPES Grenoble
  and KTH Stockholm} (\bibinfo {year} {1999})\BibitemShut {NoStop}%
\bibitem [{\citenamefont {Shafarman}\ \emph {et~al.}(1988)\citenamefont
  {Shafarman}, \citenamefont {Koon},\ and\ \citenamefont
  {Kastner}}]{CastnerPRB88}%
  \BibitemOpen
  \bibfield  {author} {\bibinfo {author} {\bibfnamefont {W.~N.}\ \bibnamefont
  {Shafarman}}, \bibinfo {author} {\bibfnamefont {D.~W.}\ \bibnamefont {Koon}},
  \ and\ \bibinfo {author} {\bibfnamefont {T.~G.}\ \bibnamefont {Kastner}},\
  }\href@noop {} {\bibfield  {journal} {\bibinfo  {journal} {Phys. Rev. B}\
  }\textbf {\bibinfo {volume} {40}},\ \bibinfo {pages} {1216} (\bibinfo {year}
  {1988})}\BibitemShut {NoStop}%
\bibitem [{\citenamefont {Delahaye}\ \emph
  {et~al.}(2003{\natexlab{b}})\citenamefont {Delahaye}, \citenamefont {Schaub},
  \citenamefont {Berger},\ and\ \citenamefont {Calvayrac}}]{DelahayePRB03}%
  \BibitemOpen
  \bibfield  {author} {\bibinfo {author} {\bibfnamefont {J.}~\bibnamefont
  {Delahaye}}, \bibinfo {author} {\bibfnamefont {T.}~\bibnamefont {Schaub}},
  \bibinfo {author} {\bibfnamefont {C.}~\bibnamefont {Berger}}, \ and\ \bibinfo
  {author} {\bibfnamefont {Y.}~\bibnamefont {Calvayrac}},\ }\href {\doibase
  10.1103/PhysRevB.67.214201} {\bibfield  {journal} {\bibinfo  {journal} {Phys.
  Rev. B}\ }\textbf {\bibinfo {volume} {67}},\ \bibinfo {pages} {214201}
  (\bibinfo {year} {2003}{\natexlab{b}})}\BibitemShut {NoStop}%
\bibitem [{\citenamefont {Delahaye}\ and\ \citenamefont
  {Berger}(2001)}]{DelahayePRB01}%
  \BibitemOpen
  \bibfield  {author} {\bibinfo {author} {\bibfnamefont {J.}~\bibnamefont
  {Delahaye}}\ and\ \bibinfo {author} {\bibfnamefont {C.}~\bibnamefont
  {Berger}},\ }\href {\doibase 10.1103/PhysRevB.64.094203} {\bibfield
  {journal} {\bibinfo  {journal} {Phys. Rev. B}\ }\textbf {\bibinfo {volume}
  {64}},\ \bibinfo {pages} {094203} (\bibinfo {year} {2001})}\BibitemShut
  {NoStop}%
\bibitem [{\citenamefont {Lanco}\ \emph {et~al.}(1992)\citenamefont {Lanco},
  \citenamefont {Klein}, \citenamefont {Berger}, \citenamefont
  {Cyrot-Lackmann}, \citenamefont {Fourcaudot},\ and\ \citenamefont
  {Sulpice}}]{LancoEPL92}%
  \BibitemOpen
  \bibfield  {author} {\bibinfo {author} {\bibfnamefont {P.}~\bibnamefont
  {Lanco}}, \bibinfo {author} {\bibfnamefont {T.}~\bibnamefont {Klein}},
  \bibinfo {author} {\bibfnamefont {C.}~\bibnamefont {Berger}}, \bibinfo
  {author} {\bibfnamefont {F.}~\bibnamefont {Cyrot-Lackmann}}, \bibinfo
  {author} {\bibfnamefont {G.}~\bibnamefont {Fourcaudot}}, \ and\ \bibinfo
  {author} {\bibfnamefont {A.}~\bibnamefont {Sulpice}},\ }\href@noop {}
  {\bibfield  {journal} {\bibinfo  {journal} {Europhys. Lett.}\ }\textbf
  {\bibinfo {volume} {18}},\ \bibinfo {pages} {227} (\bibinfo {year}
  {1992})}\BibitemShut {NoStop}%
\bibitem [{\citenamefont {Pr\'ejean}\ \emph {et~al.}(2002)\citenamefont
  {Pr\'ejean}, \citenamefont {Berger}, \citenamefont {Sulpice},\ and\
  \citenamefont {Calvayrac}}]{PrejeanPRB02}%
  \BibitemOpen
  \bibfield  {author} {\bibinfo {author} {\bibfnamefont {J.~J.}\ \bibnamefont
  {Pr\'ejean}}, \bibinfo {author} {\bibfnamefont {C.}~\bibnamefont {Berger}},
  \bibinfo {author} {\bibfnamefont {A.}~\bibnamefont {Sulpice}}, \ and\
  \bibinfo {author} {\bibfnamefont {Y.}~\bibnamefont {Calvayrac}},\ }\href
  {\doibase 10.1103/PhysRevB.65.140203} {\bibfield  {journal} {\bibinfo
  {journal} {Phys. Rev. B}\ }\textbf {\bibinfo {volume} {65}},\ \bibinfo
  {pages} {140203} (\bibinfo {year} {2002})}\BibitemShut {NoStop}%
\bibitem [{\citenamefont {Grenet}(1999)}]{GrenetAussois}%
  \BibitemOpen
  \bibfield  {author} {\bibinfo {author} {\bibfnamefont {T.}~\bibnamefont
  {Grenet}},\ }in\ \href@noop {} {\emph {\bibinfo {booktitle} {Quasicrystals
  (Proceedings of the Spring School on Quasicrystals, Aussois, 1999)}}},\
  \bibinfo {editor} {edited by\ \bibinfo {editor} {\bibfnamefont
  {E.}~\bibnamefont {Belin-Ferr\'e}}, \bibinfo {editor} {\bibfnamefont
  {C.}~\bibnamefont {Berger}}, \bibinfo {editor} {\bibfnamefont
  {M.}~\bibnamefont {Quiquandon}}, \ and\ \bibinfo {editor} {\bibfnamefont
  {A.}~\bibnamefont {Sadoc}}}\ (\bibinfo  {publisher} {World Scientific},\
  \bibinfo {year} {1999})\ p.\ \bibinfo {pages} {455}\BibitemShut {NoStop}%
\bibitem [{\citenamefont {Grenet}\ \emph {et~al.}(2007)\citenamefont {Grenet},
  \citenamefont {Delahaye}, \citenamefont {Sabra},\ and\ \citenamefont
  {Gay}}]{GrenetEPJB07}%
  \BibitemOpen
  \bibfield  {author} {\bibinfo {author} {\bibfnamefont {T.}~\bibnamefont
  {Grenet}}, \bibinfo {author} {\bibfnamefont {J.}~\bibnamefont {Delahaye}},
  \bibinfo {author} {\bibfnamefont {M.}~\bibnamefont {Sabra}}, \ and\ \bibinfo
  {author} {\bibfnamefont {F.}~\bibnamefont {Gay}},\ }\href@noop {} {\bibfield
  {journal} {\bibinfo  {journal} {Eur. Phys. J. B}\ }\textbf {\bibinfo {volume}
  {56}},\ \bibinfo {pages} {183} (\bibinfo {year} {2007})}\BibitemShut
  {NoStop}%
\bibitem [{\citenamefont {Simonet}\ \emph {et~al.}(1998)\citenamefont
  {Simonet}, \citenamefont {Hippert}, \citenamefont {Gignoux}, \citenamefont
  {Berger},\ and\ \citenamefont {Calvayrac}}]{SimonetICQ6}%
  \BibitemOpen
  \bibfield  {author} {\bibinfo {author} {\bibfnamefont {V.}~\bibnamefont
  {Simonet}}, \bibinfo {author} {\bibfnamefont {F.}~\bibnamefont {Hippert}},
  \bibinfo {author} {\bibfnamefont {C.}~\bibnamefont {Gignoux}}, \bibinfo
  {author} {\bibfnamefont {C.}~\bibnamefont {Berger}}, \ and\ \bibinfo {author}
  {\bibfnamefont {Y.}~\bibnamefont {Calvayrac}},\ }in\ \href@noop {} {\emph
  {\bibinfo {booktitle} {Proceedings of the 6th International Conference on
  Quasicrystals (ICQ6)}}},\ \bibinfo {editor} {edited by\ \bibinfo {editor}
  {\bibfnamefont {S.}~\bibnamefont {Takeuchi}}\ and\ \bibinfo {editor}
  {\bibfnamefont {T.}~\bibnamefont {Fujiwara}}}\ (\bibinfo  {publisher} {World
  Scientific},\ \bibinfo {year} {1998})\ p.\ \bibinfo {pages} {696}\BibitemShut
  {NoStop}%
\bibitem [{\citenamefont {Tang}\ \emph {et~al.}(1997)\citenamefont {Tang},
  \citenamefont {Hill}, \citenamefont {Wonnell}, \citenamefont {Poon},\ and\
  \citenamefont {Wu}}]{TangPRL97}%
  \BibitemOpen
  \bibfield  {author} {\bibinfo {author} {\bibfnamefont {X.-P.}\ \bibnamefont
  {Tang}}, \bibinfo {author} {\bibfnamefont {E.~A.}\ \bibnamefont {Hill}},
  \bibinfo {author} {\bibfnamefont {S.~K.}\ \bibnamefont {Wonnell}}, \bibinfo
  {author} {\bibfnamefont {S.~J.}\ \bibnamefont {Poon}}, \ and\ \bibinfo
  {author} {\bibfnamefont {Y.}~\bibnamefont {Wu}},\ }\href {\doibase
  10.1103/PhysRevLett.79.1070} {\bibfield  {journal} {\bibinfo  {journal}
  {Phys. Rev. Lett.}\ }\textbf {\bibinfo {volume} {79}},\ \bibinfo {pages}
  {1070} (\bibinfo {year} {1997})}\BibitemShut {NoStop}%
\bibitem [{\citenamefont {Chernikov}\ \emph {et~al.}(1996)\citenamefont
  {Chernikov}, \citenamefont {Bianchi}, \citenamefont {Felder}, \citenamefont
  {Gubler},\ and\ \citenamefont {Ott}}]{ChernikovEL96}%
  \BibitemOpen
  \bibfield  {author} {\bibinfo {author} {\bibfnamefont {M.~A.}\ \bibnamefont
  {Chernikov}}, \bibinfo {author} {\bibfnamefont {A.}~\bibnamefont {Bianchi}},
  \bibinfo {author} {\bibfnamefont {E.}~\bibnamefont {Felder}}, \bibinfo
  {author} {\bibfnamefont {U.}~\bibnamefont {Gubler}}, \ and\ \bibinfo {author}
  {\bibfnamefont {H.~R.}\ \bibnamefont {Ott}},\ }\href@noop {} {\bibfield
  {journal} {\bibinfo  {journal} {Europhys. Lett.}\ }\textbf {\bibinfo {volume}
  {35}},\ \bibinfo {pages} {431} (\bibinfo {year} {1996})}\BibitemShut
  {NoStop}%
\bibitem [{\citenamefont {Pr\'ejean}\ \emph {et~al.}(2000)\citenamefont
  {Pr\'ejean}, \citenamefont {Lasjaunias}, \citenamefont {Berger},\ and\
  \citenamefont {Sulpice}}]{PrejeanPRB00}%
  \BibitemOpen
  \bibfield  {author} {\bibinfo {author} {\bibfnamefont {J.~J.}\ \bibnamefont
  {Pr\'ejean}}, \bibinfo {author} {\bibfnamefont {J.~C.}\ \bibnamefont
  {Lasjaunias}}, \bibinfo {author} {\bibfnamefont {C.}~\bibnamefont {Berger}},
  \ and\ \bibinfo {author} {\bibfnamefont {A.}~\bibnamefont {Sulpice}},\
  }\href@noop {} {\bibfield  {journal} {\bibinfo  {journal} {Phys. Rev. B}\
  }\textbf {\bibinfo {volume} {61}},\ \bibinfo {pages} {9356} (\bibinfo {year}
  {2000})}\BibitemShut {NoStop}%
\bibitem [{\citenamefont {Davydov}\ \emph {et~al.}(1996)\citenamefont
  {Davydov}, \citenamefont {Mayou}, \citenamefont {Berger}, \citenamefont
  {Gignoux}, \citenamefont {Neumann}, \citenamefont {Jansen},\ and\
  \citenamefont {Wyder}}]{DavydovPRL96}%
  \BibitemOpen
  \bibfield  {author} {\bibinfo {author} {\bibfnamefont {D.~N.}\ \bibnamefont
  {Davydov}}, \bibinfo {author} {\bibfnamefont {D.}~\bibnamefont {Mayou}},
  \bibinfo {author} {\bibfnamefont {C.}~\bibnamefont {Berger}}, \bibinfo
  {author} {\bibfnamefont {C.}~\bibnamefont {Gignoux}}, \bibinfo {author}
  {\bibfnamefont {A.}~\bibnamefont {Neumann}}, \bibinfo {author} {\bibfnamefont
  {A.~G.~M.}\ \bibnamefont {Jansen}}, \ and\ \bibinfo {author} {\bibfnamefont
  {P.}~\bibnamefont {Wyder}},\ }\href@noop {} {\bibfield  {journal} {\bibinfo
  {journal} {Phys. Rev. Lett.}\ }\textbf {\bibinfo {volume} {77}},\ \bibinfo
  {pages} {3173} (\bibinfo {year} {1996})}\BibitemShut {NoStop}%
\bibitem [{\citenamefont {Schaub}\ \emph {et~al.}(2001)\citenamefont {Schaub},
  \citenamefont {Delahaye}, \citenamefont {Berger}, \citenamefont {Guyot},
  \citenamefont {Belkhou}, \citenamefont {Taleb-Ibrahimi},\ and\ \citenamefont
  {Calvayrac}}]{SchaubEPJB00}%
  \BibitemOpen
  \bibfield  {author} {\bibinfo {author} {\bibfnamefont {T.}~\bibnamefont
  {Schaub}}, \bibinfo {author} {\bibfnamefont {J.}~\bibnamefont {Delahaye}},
  \bibinfo {author} {\bibfnamefont {C.}~\bibnamefont {Berger}}, \bibinfo
  {author} {\bibfnamefont {H.}~\bibnamefont {Guyot}}, \bibinfo {author}
  {\bibfnamefont {R.}~\bibnamefont {Belkhou}}, \bibinfo {author} {\bibfnamefont
  {A.}~\bibnamefont {Taleb-Ibrahimi}}, \ and\ \bibinfo {author} {\bibfnamefont
  {Y.}~\bibnamefont {Calvayrac}},\ }\href@noop {} {\bibfield  {journal}
  {\bibinfo  {journal} {Eur. Phys. J. B}\ }\textbf {\bibinfo {volume} {20}},\
  \bibinfo {pages} {183} (\bibinfo {year} {2001})}\BibitemShut {NoStop}%
\bibitem [{\citenamefont {Stadnik}\ \emph {et~al.}(1996)\citenamefont
  {Stadnik}, \citenamefont {Purdie}, \citenamefont {Garnier}, \citenamefont
  {Baer}, \citenamefont {Tsai}, \citenamefont {Inoue}, \citenamefont
  {Edagawa},\ and\ \citenamefont {Takeuchi}}]{StadnikPRL96}%
  \BibitemOpen
  \bibfield  {author} {\bibinfo {author} {\bibfnamefont {Z.~M.}\ \bibnamefont
  {Stadnik}}, \bibinfo {author} {\bibfnamefont {D.}~\bibnamefont {Purdie}},
  \bibinfo {author} {\bibfnamefont {M.}~\bibnamefont {Garnier}}, \bibinfo
  {author} {\bibfnamefont {Y.}~\bibnamefont {Baer}}, \bibinfo {author}
  {\bibfnamefont {A.~P.}\ \bibnamefont {Tsai}}, \bibinfo {author}
  {\bibfnamefont {A.}~\bibnamefont {Inoue}}, \bibinfo {author} {\bibfnamefont
  {K.}~\bibnamefont {Edagawa}}, \ and\ \bibinfo {author} {\bibfnamefont
  {S.}~\bibnamefont {Takeuchi}},\ }\href {\doibase 10.1103/PhysRevLett.77.1777}
  {\bibfield  {journal} {\bibinfo  {journal} {Phys. Rev. Lett.}\ }\textbf
  {\bibinfo {volume} {77}},\ \bibinfo {pages} {1777} (\bibinfo {year}
  {1996})}\BibitemShut {NoStop}%
\bibitem [{\citenamefont {Giroud}\ \emph {et~al.}(1996)\citenamefont {Giroud},
  \citenamefont {Grenet}, \citenamefont {Linqvist}, \citenamefont {Gignoux},\
  and\ \citenamefont {Fourcaudot}}]{Giroud96}%
  \BibitemOpen
  \bibfield  {author} {\bibinfo {author} {\bibfnamefont {F.}~\bibnamefont
  {Giroud}}, \bibinfo {author} {\bibfnamefont {T.}~\bibnamefont {Grenet}},
  \bibinfo {author} {\bibfnamefont {P.}~\bibnamefont {Linqvist}}, \bibinfo
  {author} {\bibfnamefont {C.}~\bibnamefont {Gignoux}}, \ and\ \bibinfo
  {author} {\bibfnamefont {G.}~\bibnamefont {Fourcaudot}},\ }\href@noop {}
  {\bibfield  {journal} {\bibinfo  {journal} {Czechoslovak J. of Phys.}\
  }\textbf {\bibinfo {volume} {46}},\ \bibinfo {pages} {2709} (\bibinfo {year}
  {1996})},\ \bibinfo {note} {proceedings of the 21st International Conference
  on Low Temperature Physics, Prague}\BibitemShut {NoStop}%
\bibitem [{\citenamefont {Pierce}\ \emph
  {et~al.}(1993{\natexlab{a}})\citenamefont {Pierce}, \citenamefont {Bancel},
  \citenamefont {Biggs}, \citenamefont {Guo},\ and\ \citenamefont
  {Poon}}]{PiercePRB93}%
  \BibitemOpen
  \bibfield  {author} {\bibinfo {author} {\bibfnamefont {F.~S.}\ \bibnamefont
  {Pierce}}, \bibinfo {author} {\bibfnamefont {P.~A.}\ \bibnamefont {Bancel}},
  \bibinfo {author} {\bibfnamefont {B.~D.}\ \bibnamefont {Biggs}}, \bibinfo
  {author} {\bibfnamefont {Q.}~\bibnamefont {Guo}}, \ and\ \bibinfo {author}
  {\bibfnamefont {S.~J.}\ \bibnamefont {Poon}},\ }\href {\doibase
  10.1103/PhysRevB.47.5670} {\bibfield  {journal} {\bibinfo  {journal} {Phys.
  Rev. B}\ }\textbf {\bibinfo {volume} {47}},\ \bibinfo {pages} {5670}
  (\bibinfo {year} {1993}{\natexlab{a}})}\BibitemShut {NoStop}%
\bibitem [{\citenamefont {Delahaye}\ \emph {et~al.}(1999)\citenamefont
  {Delahaye}, \citenamefont {Gignoux}, \citenamefont {Schaub}, \citenamefont
  {Berger}, \citenamefont {Grenet}, \citenamefont {Sulpice}, \citenamefont
  {Pr\'ejean},\ and\ \citenamefont {Lasjaunias}}]{DelahayeJNCS99}%
  \BibitemOpen
  \bibfield  {author} {\bibinfo {author} {\bibfnamefont {J.}~\bibnamefont
  {Delahaye}}, \bibinfo {author} {\bibfnamefont {C.}~\bibnamefont {Gignoux}},
  \bibinfo {author} {\bibfnamefont {T.}~\bibnamefont {Schaub}}, \bibinfo
  {author} {\bibfnamefont {C.}~\bibnamefont {Berger}}, \bibinfo {author}
  {\bibfnamefont {T.}~\bibnamefont {Grenet}}, \bibinfo {author} {\bibfnamefont
  {A.}~\bibnamefont {Sulpice}}, \bibinfo {author} {\bibfnamefont {J.~J.}\
  \bibnamefont {Pr\'ejean}}, \ and\ \bibinfo {author} {\bibfnamefont {J.~C.}\
  \bibnamefont {Lasjaunias}},\ }\href@noop {} {\bibfield  {journal} {\bibinfo
  {journal} {J. of Non Cryst. Sol.}\ }\textbf {\bibinfo {volume} {250-252}},\
  \bibinfo {pages} {878} (\bibinfo {year} {1999})}\BibitemShut {NoStop}%
\bibitem [{\citenamefont {Rodmar}\ \emph
  {et~al.}(1999{\natexlab{a}})\citenamefont {Rodmar}, \citenamefont
  {Oberschmidt}, \citenamefont {Ahlgren}, \citenamefont {Gignoux},
  \citenamefont {Delahaye}, \citenamefont {Berger}, \citenamefont {Poon},\ and\
  \citenamefont {Rapp}}]{RodmarJNCS99}%
  \BibitemOpen
  \bibfield  {author} {\bibinfo {author} {\bibfnamefont {M.}~\bibnamefont
  {Rodmar}}, \bibinfo {author} {\bibfnamefont {D.}~\bibnamefont {Oberschmidt}},
  \bibinfo {author} {\bibfnamefont {M.}~\bibnamefont {Ahlgren}}, \bibinfo
  {author} {\bibfnamefont {C.}~\bibnamefont {Gignoux}}, \bibinfo {author}
  {\bibfnamefont {J.}~\bibnamefont {Delahaye}}, \bibinfo {author}
  {\bibfnamefont {C.}~\bibnamefont {Berger}}, \bibinfo {author} {\bibfnamefont
  {S.~J.}\ \bibnamefont {Poon}}, \ and\ \bibinfo {author} {\bibfnamefont
  {O.}~\bibnamefont {Rapp}},\ }\href@noop {} {\bibfield  {journal} {\bibinfo
  {journal} {J. of Non Cryst. Sol.}\ }\textbf {\bibinfo {volume} {250-252}},\
  \bibinfo {pages} {883} (\bibinfo {year} {1999}{\natexlab{a}})}\BibitemShut
  {NoStop}%
\bibitem [{\citenamefont {Rodmar}\ \emph {et~al.}(2000)\citenamefont {Rodmar},
  \citenamefont {Ahlgren}, \citenamefont {Oberschmidt}, \citenamefont
  {Gignoux}, \citenamefont {Delahaye}, \citenamefont {Berger}, \citenamefont
  {Poon},\ and\ \citenamefont {Rapp}}]{RodmarPRB00}%
  \BibitemOpen
  \bibfield  {author} {\bibinfo {author} {\bibfnamefont {M.}~\bibnamefont
  {Rodmar}}, \bibinfo {author} {\bibfnamefont {M.}~\bibnamefont {Ahlgren}},
  \bibinfo {author} {\bibfnamefont {D.}~\bibnamefont {Oberschmidt}}, \bibinfo
  {author} {\bibfnamefont {C.}~\bibnamefont {Gignoux}}, \bibinfo {author}
  {\bibfnamefont {J.}~\bibnamefont {Delahaye}}, \bibinfo {author}
  {\bibfnamefont {C.}~\bibnamefont {Berger}}, \bibinfo {author} {\bibfnamefont
  {S.~J.}\ \bibnamefont {Poon}}, \ and\ \bibinfo {author} {\bibfnamefont
  {O.}~\bibnamefont {Rapp}},\ }\href {\doibase 10.1103/PhysRevB.61.3936}
  {\bibfield  {journal} {\bibinfo  {journal} {Phys. Rev. B}\ }\textbf {\bibinfo
  {volume} {61}},\ \bibinfo {pages} {3936} (\bibinfo {year}
  {2000})}\BibitemShut {NoStop}%
\bibitem [{\citenamefont {Gignoux}(1996)}]{GignouxThesis96}%
  \BibitemOpen
  \bibfield  {author} {\bibinfo {author} {\bibfnamefont {C.}~\bibnamefont
  {Gignoux}},\ }\emph {\bibinfo {title} {Etude des Propri\'et\'es Electroniques
  de l'Alliage Quasicristallin AlPdRe}},\ \href@noop {} {\bibinfo {type}
  {{Ph.D.} thesis}},\ \bibinfo  {school} {Universit\'e Joseph Fourier Grenoble}
  (\bibinfo {year} {1996})\BibitemShut {NoStop}%
\bibitem [{\citenamefont {Kirihara}\ and\ \citenamefont
  {Kimura}(1998)}]{KiriharaMRS99}%
  \BibitemOpen
  \bibfield  {author} {\bibinfo {author} {\bibfnamefont {K.}~\bibnamefont
  {Kirihara}}\ and\ \bibinfo {author} {\bibfnamefont {K.}~\bibnamefont
  {Kimura}},\ }\href {\doibase 10.1557/PROC-553-379} {\bibfield  {journal}
  {\bibinfo  {journal} {MRS Proceedings}\ }\textbf {\bibinfo {volume} {553}},\
  \bibinfo {pages} {379} (\bibinfo {year} {1998})}\BibitemShut {NoStop}%
\bibitem [{\citenamefont {Tsai}\ \emph {et~al.}(1990)\citenamefont {Tsai},
  \citenamefont {Inoue}, \citenamefont {Yokoyama},\ and\ \citenamefont
  {Masumoto}}]{TsaiMTJIM90}%
  \BibitemOpen
  \bibfield  {author} {\bibinfo {author} {\bibfnamefont {A.~P.}\ \bibnamefont
  {Tsai}}, \bibinfo {author} {\bibfnamefont {A.}~\bibnamefont {Inoue}},
  \bibinfo {author} {\bibfnamefont {Y.}~\bibnamefont {Yokoyama}}, \ and\
  \bibinfo {author} {\bibfnamefont {T.}~\bibnamefont {Masumoto}},\ }\href@noop
  {} {\bibfield  {journal} {\bibinfo  {journal} {Materials Transactions JIM}\
  }\textbf {\bibinfo {volume} {31}},\ \bibinfo {pages} {98} (\bibinfo {year}
  {1990})}\BibitemShut {NoStop}%
\bibitem [{\citenamefont {Pierce}\ \emph
  {et~al.}(1993{\natexlab{b}})\citenamefont {Pierce}, \citenamefont {Poon},\
  and\ \citenamefont {Guo}}]{PierceScience93}%
  \BibitemOpen
  \bibfield  {author} {\bibinfo {author} {\bibfnamefont {F.~S.}\ \bibnamefont
  {Pierce}}, \bibinfo {author} {\bibfnamefont {S.~J.}\ \bibnamefont {Poon}}, \
  and\ \bibinfo {author} {\bibfnamefont {Q.}~\bibnamefont {Guo}},\ }\href@noop
  {} {\bibfield  {journal} {\bibinfo  {journal} {Science}\ }\textbf {\bibinfo
  {volume} {261}},\ \bibinfo {pages} {737} (\bibinfo {year}
  {1993}{\natexlab{b}})}\BibitemShut {NoStop}%
\bibitem [{\citenamefont {Lin}\ \emph {et~al.}(1997)\citenamefont {Lin},
  \citenamefont {Lin}, \citenamefont {Wang}, \citenamefont {Chou},
  \citenamefont {Horng}, \citenamefont {Cheng}, \citenamefont {Yao},\ and\
  \citenamefont {Lai}}]{LinJPCM97}%
  \BibitemOpen
  \bibfield  {author} {\bibinfo {author} {\bibfnamefont {C.~R.}\ \bibnamefont
  {Lin}}, \bibinfo {author} {\bibfnamefont {S.~T.}\ \bibnamefont {Lin}},
  \bibinfo {author} {\bibfnamefont {C.~W.}\ \bibnamefont {Wang}}, \bibinfo
  {author} {\bibfnamefont {S.~L.}\ \bibnamefont {Chou}}, \bibinfo {author}
  {\bibfnamefont {H.~E.}\ \bibnamefont {Horng}}, \bibinfo {author}
  {\bibfnamefont {J.~M.}\ \bibnamefont {Cheng}}, \bibinfo {author}
  {\bibfnamefont {Y.~D.}\ \bibnamefont {Yao}}, \ and\ \bibinfo {author}
  {\bibfnamefont {S.~C.}\ \bibnamefont {Lai}},\ }\href@noop {} {\bibfield
  {journal} {\bibinfo  {journal} {J. Phys.: Condens. Matter}\ }\textbf
  {\bibinfo {volume} {9}},\ \bibinfo {pages} {1509} (\bibinfo {year}
  {1997})}\BibitemShut {NoStop}%
\bibitem [{\citenamefont {Lindqvist}\ \emph {et~al.}(1993)\citenamefont
  {Lindqvist}, \citenamefont {Berger}, \citenamefont {Klein}, \citenamefont
  {Lanco}, \citenamefont {Cyrot-Lackmann},\ and\ \citenamefont
  {Calvayrac}}]{LindqvistPRB93}%
  \BibitemOpen
  \bibfield  {author} {\bibinfo {author} {\bibfnamefont {P.}~\bibnamefont
  {Lindqvist}}, \bibinfo {author} {\bibfnamefont {C.}~\bibnamefont {Berger}},
  \bibinfo {author} {\bibfnamefont {T.}~\bibnamefont {Klein}}, \bibinfo
  {author} {\bibfnamefont {P.}~\bibnamefont {Lanco}}, \bibinfo {author}
  {\bibfnamefont {F.}~\bibnamefont {Cyrot-Lackmann}}, \ and\ \bibinfo {author}
  {\bibfnamefont {Y.}~\bibnamefont {Calvayrac}},\ }\href {\doibase
  10.1103/PhysRevB.48.630} {\bibfield  {journal} {\bibinfo  {journal} {Phys.
  Rev. B}\ }\textbf {\bibinfo {volume} {48}},\ \bibinfo {pages} {630} (\bibinfo
  {year} {1993})}\BibitemShut {NoStop}%
\bibitem [{\citenamefont {Mayou}\ \emph {et~al.}(1993)\citenamefont {Mayou},
  \citenamefont {Berger}, \citenamefont {Cyrot-Lackmann}, \citenamefont
  {Klein},\ and\ \citenamefont {Lanco}}]{MayouPRL93}%
  \BibitemOpen
  \bibfield  {author} {\bibinfo {author} {\bibfnamefont {D.}~\bibnamefont
  {Mayou}}, \bibinfo {author} {\bibfnamefont {C.}~\bibnamefont {Berger}},
  \bibinfo {author} {\bibfnamefont {F.}~\bibnamefont {Cyrot-Lackmann}},
  \bibinfo {author} {\bibfnamefont {T.}~\bibnamefont {Klein}}, \ and\ \bibinfo
  {author} {\bibfnamefont {P.}~\bibnamefont {Lanco}},\ }\href {\doibase
  10.1103/PhysRevLett.70.3915} {\bibfield  {journal} {\bibinfo  {journal}
  {Phys. Rev. Lett.}\ }\textbf {\bibinfo {volume} {70}},\ \bibinfo {pages}
  {3915} (\bibinfo {year} {1993})}\BibitemShut {NoStop}%
\bibitem [{\citenamefont {Haberkern}\ \emph {et~al.}(2000)\citenamefont
  {Haberkern}, \citenamefont {Khedhri}, \citenamefont {Madel},\ and\
  \citenamefont {H\"aussler}}]{HaberkernICQ7}%
  \BibitemOpen
  \bibfield  {author} {\bibinfo {author} {\bibfnamefont {R.}~\bibnamefont
  {Haberkern}}, \bibinfo {author} {\bibfnamefont {K.}~\bibnamefont {Khedhri}},
  \bibinfo {author} {\bibfnamefont {C.}~\bibnamefont {Madel}}, \ and\ \bibinfo
  {author} {\bibfnamefont {P.}~\bibnamefont {H\"aussler}},\ }\href@noop {}
  {\bibfield  {journal} {\bibinfo  {journal} {Materials Science and Engineering
  A}\ }\textbf {\bibinfo {volume} {294-296}},\ \bibinfo {pages} {475} (\bibinfo
  {year} {2000})}\BibitemShut {NoStop}%
\bibitem [{\citenamefont {Tsai}(1999)}]{TsaiBook99}%
  \BibitemOpen
  \bibfield  {author} {\bibinfo {author} {\bibfnamefont {A.~P.}\ \bibnamefont
  {Tsai}},\ }\enquote {\bibinfo {title} {Physical properties of
  quasicrystals},}\ \ (\bibinfo  {publisher} {Springer},\ \bibinfo {year}
  {1999})\ Chap.~\bibinfo {chapter} {2}, pp.\ \bibinfo {pages}
  {5--50}\BibitemShut {NoStop}%
\bibitem [{\citenamefont {Karkin}\ \emph {et~al.}(2002)\citenamefont {Karkin},
  \citenamefont {Goshchitskii}, \citenamefont {Voronin}, \citenamefont {Poon},
  \citenamefont {Srinivas},\ and\ \citenamefont {Rapp}}]{KarkinPRB02}%
  \BibitemOpen
  \bibfield  {author} {\bibinfo {author} {\bibfnamefont {A.~E.}\ \bibnamefont
  {Karkin}}, \bibinfo {author} {\bibfnamefont {B.~N.}\ \bibnamefont
  {Goshchitskii}}, \bibinfo {author} {\bibfnamefont {V.~I.}\ \bibnamefont
  {Voronin}}, \bibinfo {author} {\bibfnamefont {S.~J.}\ \bibnamefont {Poon}},
  \bibinfo {author} {\bibfnamefont {V.}~\bibnamefont {Srinivas}}, \ and\
  \bibinfo {author} {\bibfnamefont {O.}~\bibnamefont {Rapp}},\ }\href@noop {}
  {\bibfield  {journal} {\bibinfo  {journal} {Phys. Rev. B}\ }\textbf {\bibinfo
  {volume} {66}},\ \bibinfo {pages} {092203} (\bibinfo {year}
  {2002})}\BibitemShut {NoStop}%
\bibitem [{\citenamefont {Rapp}\ \emph {et~al.}(2008)\citenamefont {Rapp},
  \citenamefont {Karkin}, \citenamefont {Goshchitskii}, \citenamefont
  {Voronin}, \citenamefont {Srinivas},\ and\ \citenamefont
  {Poon}}]{RappJPCM08}%
  \BibitemOpen
  \bibfield  {author} {\bibinfo {author} {\bibfnamefont {O.}~\bibnamefont
  {Rapp}}, \bibinfo {author} {\bibfnamefont {A.~A.}\ \bibnamefont {Karkin}},
  \bibinfo {author} {\bibfnamefont {B.~N.}\ \bibnamefont {Goshchitskii}},
  \bibinfo {author} {\bibfnamefont {V.~I.}\ \bibnamefont {Voronin}}, \bibinfo
  {author} {\bibfnamefont {V.}~\bibnamefont {Srinivas}}, \ and\ \bibinfo
  {author} {\bibfnamefont {S.~J.}\ \bibnamefont {Poon}},\ }\href@noop {}
  {\bibfield  {journal} {\bibinfo  {journal} {J. Phys.: Condens. Matter}\
  }\textbf {\bibinfo {volume} {20}},\ \bibinfo {pages} {114120} (\bibinfo
  {year} {2008})}\BibitemShut {NoStop}%
\bibitem [{\citenamefont {Beeli}\ \emph {et~al.}(2000)\citenamefont {Beeli},
  \citenamefont {Soltmann},\ and\ \citenamefont {Poon}}]{BeeliICQ7}%
  \BibitemOpen
  \bibfield  {author} {\bibinfo {author} {\bibfnamefont {C.}~\bibnamefont
  {Beeli}}, \bibinfo {author} {\bibfnamefont {C.}~\bibnamefont {Soltmann}}, \
  and\ \bibinfo {author} {\bibfnamefont {S.~J.}\ \bibnamefont {Poon}},\
  }\href@noop {} {\bibfield  {journal} {\bibinfo  {journal} {Materials Science
  and Engineering A}\ }\textbf {\bibinfo {volume} {294-296}},\ \bibinfo {pages}
  {531} (\bibinfo {year} {2000})}\BibitemShut {NoStop}%
\bibitem [{\citenamefont {Guo}\ and\ \citenamefont {Poon}(1996)}]{GuoPRB96}%
  \BibitemOpen
  \bibfield  {author} {\bibinfo {author} {\bibfnamefont {Q.}~\bibnamefont
  {Guo}}\ and\ \bibinfo {author} {\bibfnamefont {S.~J.}\ \bibnamefont {Poon}},\
  }\href@noop {} {\bibfield  {journal} {\bibinfo  {journal} {Phys. Rev. B}\
  }\textbf {\bibinfo {volume} {54}},\ \bibinfo {pages} {12793} (\bibinfo {year}
  {1996})}\BibitemShut {NoStop}%
\bibitem [{\citenamefont {Lin}\ \emph {et~al.}(1996)\citenamefont {Lin},
  \citenamefont {Chou},\ and\ \citenamefont {Lin}}]{LinJPCM96}%
  \BibitemOpen
  \bibfield  {author} {\bibinfo {author} {\bibfnamefont {C.~R.}\ \bibnamefont
  {Lin}}, \bibinfo {author} {\bibfnamefont {S.~L.}\ \bibnamefont {Chou}}, \
  and\ \bibinfo {author} {\bibfnamefont {S.~T.}\ \bibnamefont {Lin}},\
  }\href@noop {} {\bibfield  {journal} {\bibinfo  {journal} {J. Phys.: Condens.
  Matter}\ }\textbf {\bibinfo {volume} {8}},\ \bibinfo {pages} {L725} (\bibinfo
  {year} {1996})}\BibitemShut {NoStop}%
\bibitem [{\citenamefont {Wang}\ \emph {et~al.}(1998)\citenamefont {Wang},
  \citenamefont {Su},\ and\ \citenamefont {Lin}}]{WangSSC98}%
  \BibitemOpen
  \bibfield  {author} {\bibinfo {author} {\bibfnamefont {C.~R.}\ \bibnamefont
  {Wang}}, \bibinfo {author} {\bibfnamefont {Z.~Y.}\ \bibnamefont {Su}}, \ and\
  \bibinfo {author} {\bibfnamefont {S.~T.}\ \bibnamefont {Lin}},\ }\href@noop
  {} {\bibfield  {journal} {\bibinfo  {journal} {Solid State Communications}\
  }\textbf {\bibinfo {volume} {108}},\ \bibinfo {pages} {681} (\bibinfo {year}
  {1998})}\BibitemShut {NoStop}%
\bibitem [{\citenamefont {Delahaye}\ \emph {et~al.}(1998)\citenamefont
  {Delahaye}, \citenamefont {Brison},\ and\ \citenamefont
  {Berger}}]{DelahayePRL98}%
  \BibitemOpen
  \bibfield  {author} {\bibinfo {author} {\bibfnamefont {J.}~\bibnamefont
  {Delahaye}}, \bibinfo {author} {\bibfnamefont {J.~P.}\ \bibnamefont
  {Brison}}, \ and\ \bibinfo {author} {\bibfnamefont {C.}~\bibnamefont
  {Berger}},\ }\href {\doibase 10.1103/PhysRevLett.81.4204} {\bibfield
  {journal} {\bibinfo  {journal} {Phys. Rev. Lett.}\ }\textbf {\bibinfo
  {volume} {81}},\ \bibinfo {pages} {4204} (\bibinfo {year}
  {1998})}\BibitemShut {NoStop}%
\bibitem [{\citenamefont {M\"obius}\ \emph {et~al.}(1999)\citenamefont
  {M\"obius}, \citenamefont {Frenzel}, \citenamefont {Thielsch}, \citenamefont
  {Rosenbaum}, \citenamefont {Adkins}, \citenamefont {Schreiber}, \citenamefont
  {Bauer}, \citenamefont {Gr\"otzschel}, \citenamefont {Hoffmann},
  \citenamefont {Krieg}, \citenamefont {Matz}, \citenamefont {Vinzelberg},\
  and\ \citenamefont {Witcomb}}]{MobiusPRB99}%
  \BibitemOpen
  \bibfield  {author} {\bibinfo {author} {\bibfnamefont {A.}~\bibnamefont
  {M\"obius}}, \bibinfo {author} {\bibfnamefont {C.}~\bibnamefont {Frenzel}},
  \bibinfo {author} {\bibfnamefont {R.}~\bibnamefont {Thielsch}}, \bibinfo
  {author} {\bibfnamefont {R.}~\bibnamefont {Rosenbaum}}, \bibinfo {author}
  {\bibfnamefont {C.~J.}\ \bibnamefont {Adkins}}, \bibinfo {author}
  {\bibfnamefont {M.}~\bibnamefont {Schreiber}}, \bibinfo {author}
  {\bibfnamefont {H.-D.}\ \bibnamefont {Bauer}}, \bibinfo {author}
  {\bibfnamefont {R.}~\bibnamefont {Gr\"otzschel}}, \bibinfo {author}
  {\bibfnamefont {V.}~\bibnamefont {Hoffmann}}, \bibinfo {author}
  {\bibfnamefont {T.}~\bibnamefont {Krieg}}, \bibinfo {author} {\bibfnamefont
  {N.}~\bibnamefont {Matz}}, \bibinfo {author} {\bibfnamefont {H.}~\bibnamefont
  {Vinzelberg}}, \ and\ \bibinfo {author} {\bibfnamefont {M.}~\bibnamefont
  {Witcomb}},\ }\href {\doibase 10.1103/PhysRevB.60.14209} {\bibfield
  {journal} {\bibinfo  {journal} {Phys. Rev. B}\ }\textbf {\bibinfo {volume}
  {60}},\ \bibinfo {pages} {14209} (\bibinfo {year} {1999})}\BibitemShut
  {NoStop}%
\bibitem [{\citenamefont {Rodmar}\ \emph
  {et~al.}(1999{\natexlab{b}})\citenamefont {Rodmar}, \citenamefont
  {Zavaliche}, \citenamefont {Poon},\ and\ \citenamefont {Rapp}}]{RodmarPRB02}%
  \BibitemOpen
  \bibfield  {author} {\bibinfo {author} {\bibfnamefont {M.}~\bibnamefont
  {Rodmar}}, \bibinfo {author} {\bibfnamefont {F.}~\bibnamefont {Zavaliche}},
  \bibinfo {author} {\bibfnamefont {S.~J.}\ \bibnamefont {Poon}}, \ and\
  \bibinfo {author} {\bibfnamefont {O.}~\bibnamefont {Rapp}},\ }\href {\doibase
  10.1103/PhysRevB.60.10807} {\bibfield  {journal} {\bibinfo  {journal} {Phys.
  Rev. B}\ }\textbf {\bibinfo {volume} {60}},\ \bibinfo {pages} {10807}
  (\bibinfo {year} {1999}{\natexlab{b}})}\BibitemShut {NoStop}%
\bibitem [{\citenamefont {Calvayrac}()}]{CalvayracComPrive}%
  \BibitemOpen
  \bibfield  {author} {\bibinfo {author} {\bibfnamefont {Y.}~\bibnamefont
  {Calvayrac}},\ }\href@noop {} {\bibinfo  {journal} {Private communication}\
  }\BibitemShut {NoStop}%
\bibitem [{\citenamefont {Ahlgren}\ \emph {et~al.}(1997)\citenamefont
  {Ahlgren}, \citenamefont {Gignoux}, \citenamefont {Rodmar}, \citenamefont
  {Berger},\ and\ \citenamefont {Rapp}}]{AhlgrenPRB97}%
  \BibitemOpen
\bibfield  {journal} {  }\bibfield  {author} {\bibinfo {author} {\bibfnamefont
  {M.}~\bibnamefont {Ahlgren}}, \bibinfo {author} {\bibfnamefont
  {C.}~\bibnamefont {Gignoux}}, \bibinfo {author} {\bibfnamefont
  {M.}~\bibnamefont {Rodmar}}, \bibinfo {author} {\bibfnamefont
  {C.}~\bibnamefont {Berger}}, \ and\ \bibinfo {author} {\bibfnamefont
  {O.}~\bibnamefont {Rapp}},\ }\href {\doibase 10.1103/PhysRevB.55.R11915}
  {\bibfield  {journal} {\bibinfo  {journal} {Phys. Rev. B}\ }\textbf {\bibinfo
  {volume} {55}},\ \bibinfo {pages} {R11915} (\bibinfo {year}
  {1997})}\BibitemShut {NoStop}%
\end{thebibliography}
\end{document}